\documentclass[a4paper,12pt,twoside]{article}
\usepackage[german,english]{babel}
\usepackage{rotating}
\usepackage{epsfig}
\usepackage[centertags]{amsmath}
\usepackage{amssymb,amscd}
\usepackage{array}
\usepackage{multirow}
\usepackage{supertabular}
\usepackage{dcolumn}
\usepackage{./mcite}
\usepackage{cite}
\usepackage{xspace}
\def\zero{{\scriptscriptstyle 0}}
\def\sone{{\scriptscriptstyle 1}}
\def\stwo{{\scriptscriptstyle 2}}
\def\sthr{{\scriptscriptstyle 3}}

\def\none{---}

\def\Z0{{Z^\zero}}
\def\eVdist{\kern-0.06667em}

\def\Gev{{\text{Ge}\eVdist\text{V\/}}}

\def\Tev{{\text{Te}\eVdist\text{V\/}}}

\def\gev{{\,\text{Ge}\eVdist\text{V\/}}}

\def\tev{{\,\text{Te}\eVdist\text{V\/}}}

\def\pbi{\,\text{pb}^{-1}}

\def\cm{\,\text{cm}}

\def\DO{{D{\O}}\xspace}

\def\IP{{\rm I$\kern-0.01667em$P}}

\def\L{{\scriptscriptstyle{\rm L}}}

\def\NC{{\scriptscriptstyle{\rm NC}}}

\def\R{{\scriptscriptstyle{\rm R}}}
\def\SM{{\scriptscriptstyle{\rm SM}}}
\def\SU2U1{{\rm SU}(2)\times{\rm U}(1)}

\def\alsmz{{\alpha_s(M_Z^2)}}

\def\als{{\alpha_s}}

\def\ctws{\cos^2\theta_{\scriptscriptstyle W}}

\def\epslam{{\epsilon/\Lambda^2}}

\def\exp{{\rm exp}}

\def\lum{{\cal L}}

\def\slim{{\rm lim}}
\def\slimb{{\rm lim,B}}
\def\slimr{{\rm lim,R}}
\def\smin{{\rm min}}
\def\smax{{\rm max}}

\def\stws{\sin^2\theta_{\scriptscriptstyle W}}

\def\tot{{\rm tot}}
\def\true{{\rm true}}
\def\tw{\theta_{\scriptscriptstyle W}}
\mathchardef\qsm=63
\mathchardef\pls=43
\mathchardef\mns=512
\mathchardef\plm=518
\mathchardef\eql=61
\mathchardef\smallleft=300
\mathchardef\smallright=301
\mathchardef\perslsh=47
\mathchardef\les=316
\mathchardef\gre=318
\mathchardef\leq=532
\mathchardef\grq=533
\chardef\usc=95
\chardef\til=126
\def\intl{\int\limits}

\def\sqr#1#2#3{{\vcenter{\hrule height.#3ex\hbox{\vrule width.#2ex height#1ex
    \kern#1ex\vrule width.#3ex}\hrule height.#2ex}}}

\def\angleto{\vrule width.035em height2.1ex depth-.56ex\unskip\kern-.6ex\to}
\def\perchc#1{{\raise.4ex\hbox{$\mkern4mu#1{\it\perslsh}_
             {\mkern-5mu\scriptscriptstyle{{\rm o}\!{\rm o}}}^
             {\mkern-12.8mu\scriptscriptstyle{\rm o}}$}}}

\def\widebar#1{\mkern1.5mu\overline{\mkern-1.5mu#1\mkern-1.mu}\mkern1.mu}
\catcode`\@=11 
\def\parenbar{\mathpalette\p@renb@r}
\def\p@renb@r#1#2{\vbox{%
  \ifx#1\scriptscriptstyle \dimen@.7em\dimen@ii.2em\else
  \ifx#1\scriptstyle \dimen@.8em\dimen@ii.25em\else
  \dimen@1em\dimen@ii.4em\fi\fi \offinterlineskip
  \ialign{\hfill##\hfill\cr
    \vbox{\hrule width\dimen@ii}\cr
    \noalign{\vskip-.3ex}%
    \hbox to\dimen@{$\mathchar300\hfil\mathchar301$}\cr
    \noalign{\vskip-.3ex}%
    $#1#2$\cr}}}
\catcode`\@=12 

\def\pbar{\widebar{p}}
\def\qbar{\widebar{q}}

\def\nubar{\widebar{\nu}}

\def\ebar{\widebar{e}}
\newbox\struttbox
\setbox\struttbox=\hbox{\vrule height1.65ex depth.485ex width0pt}
\def\strutt{\relax\ifmmode\copy\struttbox\else\unhcopy\struttbox\fi}
\def\stru#1#2{\relax\ifmmode\hbox{\vrule height#1 depth#2 width0pt}
\else\vrule height#1 depth#2 width0pt\fi}
\def\uline#1{$\underline{\hbox{#1\strutt}}$}
\def\ronum#1{\uppercase\expandafter{\romannumeral#1}}
\def\ronuml#1{\expandafter{\romannumeral#1}}

\def\cB{{\phantom{00}}}

\def\eq#1{eq.~(\ref{eq-#1})}

\def\eqsto#1#2{eqs.~(\ref{eq-#1}-\ref{eq-#2})}

\def\fig#1{Fig.~\ref{fig-#1}}

\def\figand#1#2{Figs.~\ref{fig-#1} and~\ref{fig-#2}}
\def\figand#1#2{Figs.~\ref{fig-#1} and~\ref{fig-#2}}
\def\tab#1{Table~\ref{tab-#1}}

\def\Tab#1{Table~\ref{tab-#1}}
\def\sec#1{Sect.~\ref{sec-#1}}

\DeclareMathAlphabet{\mathbf}{OT1}{cmr}{bx}{sl}
\def\mb{\mathbf}

\renewcommand{\arraystretch}{1.1}
\newenvironment{tightitemize}[2]
 {\begin{list}{$\bullet$}%
  {\setlength{\leftmargin}{\leftmargini}%
   \setlength{\topsep}{#1}%
   \setlength{\itemsep}{#2}%
  }%
 }%
 {\end{list}}
\newenvironment{tightsubitemize}[2]
 {\begin{list}{$-$}%
  {\setlength{\leftmargin}{\leftmargini}%
   \setlength{\topsep}{#1}%
   \setlength{\itemsep}{#2}%
  }%
 }%
 {\end{list}}
\newenvironment{tightssubitemize}[2]
 {\begin{list}{$\to$}%
  {\setlength{\leftmargin}{\leftmargini}%
   \setlength{\topsep}{#1}%
   \setlength{\itemsep}{#2}%
  }%
 }%
 {\end{list}}

\catcode`\@=11 
\newlength{\@fninsert}
\newlength{\@fnwidth}
\renewcommand{\@makefntext}[1]%
  {\noindent\makebox[\@fninsert][r]{\@makefnmark}\hfil%
  \parbox[t]{\@fnwidth}{#1}}
\catcode`\@=12 
\newlength{\localtextwidth}
\catcode`\@=11 
\newsavebox{\tmpbox}
\newlength{\@captionmargin}
\newlength{\@captionwidth}
\newlength{\@captionitemtextsep}
\renewcommand{\@makecaption}[2]%
  {%
   \vspace{10.pt}
   \setlength{\@captionwidth}{\localtextwidth}
   \addtolength{\@captionwidth}{-\@captionmargin}
   \sbox{\tmpbox}{{\bf #1:}{\it #2}}%
   \ifthenelse{\lengthtest{\wd\tmpbox > \@captionwidth}}%
   {\centerline{\parbox[t]{\@captionwidth}%
   {\tolerance=2000\normalsize%
    {\bf #1:}\hspace{\@captionitemtextsep}{\it #2}}}}%
   {\centerline{{\bf #1:}\kern1.em{\it #2}}}}
\catcode`\@=12 
\catcode`\@=11 
\renewcommand\section{\@startsection{section}{1}{\z@}%
                                   {-3.5ex \@plus -1ex \@minus -.2ex}%
                                   {2.3ex \@plus.2ex}%
                                   {\normalfont\Large\bfseries}}
\renewcommand\subsection{\@startsection{subsection}{2}{\z@}%
                                   {-3.25ex\@plus -1ex \@minus -.2ex}%
                                   {1.5ex \@plus .2ex}%
                                   {\normalfont\large\bfseries}}
\renewcommand\subsubsection{\@startsection{subsubsection}{3}{\z@}%
                                   {-3.25ex\@plus -1ex \@minus -.2ex}%
                                   {1.5ex \@plus .2ex}%
                                   {\normalfont\large\bfseries}}
\renewcommand\paragraph{\@startsection{paragraph}{4}{\z@}%
                                   {3.25ex \@plus1ex \@minus.2ex}%
                                   {1.2ex \@plus .2ex}%
                                   {\normalfont\normalsize\bfseries}}
\catcode`\@=12 

\def\citeCIphe{{\cite{%
alt-97-01,bar-97-03,bar-98-01,dib-97-01%
}}\xspace}

\def\citeCILthree{{\cite{%
acc-98-01,l3-ci-van,*l3-ci-mor,*l3-ci-dim%
}}\xspace}
\def\citeCILthreecomp{{\cite{%
l3-ci-van,*l3-ci-mor,*l3-ci-dim%
}}\xspace}
\def\citeAPVlim{{\cite{%
bar-98-01,dea-97-01,giu-97-02%
}}\xspace}
\def\citeCTD{{\cite{%
zeus-det-89-04,*zeus-det-92-11,*zeus-det-94-02%
}}\xspace}
\def\citeCAL{{\cite{%
zeus-det-91-07,*zeus-det-91-08,*zeus-det-92-07,*zeus-det-93-05%
}}\xspace}
\def\citeJETSET{{\cite{%
sjo-86-01,*sjo-87-01,*sjo-94-01%
}}\xspace}

\begin{document}
\selectlanguage{english}
\thispagestyle{empty}

\title{
\rightline{\large DESY--99--058}
\vskip2.2cm
\bf\LARGE Search for Contact Interactions\\
\bf\LARGE in Deep Inelastic $\mb{e^+p\to e^+X}$ Scattering\\
\bf\LARGE at HERA
\vskip2cm\strut
}                                                       
                    
\author{ZEUS Collaboration}
\date{ }
\maketitle

\vskip2.5cm
\centerline{\bf Abstract}
\vskip4.mm
\centerline{\begin{minipage}{15.cm}
\noindent
In a search for signatures of physics processes beyond the Standard Model,
various $eeqq$ vector contact--interaction hypotheses have been tested using the
high--$Q^2$, deep inelastic neutral--current $e^+p$ scattering data collected
with the ZEUS detector at HERA.  The data correspond to an integrated luminosity
of $47.7\pbi$ of $e^+p$ interactions at $300\gev$ center--of--mass energy. No
significant evidence of a contact--interaction signal has been found.  Limits at
the $95\%$ confidence level are set on the contact--interaction amplitudes. The
effective mass scales $\Lambda$ corresponding to these limits range from
$1.7\tev$ to $5\tev$ for the contact--interaction scenarios considered.
\end{minipage}}

\thispagestyle{empty}
\newpage

%
%
%
%
{ 
\textheight 680pt
\topmargin-1.cm
\parindent0.cm 
\parskip0.3cm plus0.05cm minus0.05cm 
\def\3{\ss} 
\pagenumbering{Roman} 
                                                   %
\begin{center} 
{                      \Large  The ZEUS Collaboration              } 
\end{center} 

{\raggedright
  J.~Breitweg, 
  S.~Chekanov, 
  M.~Derrick, 
  D.~Krakauer, 
  S.~Magill, 
  B.~Musgrave, 
  A.~Pellegrino, 
  J.~Repond, 
  R.~Stanek, 
  R.~Yoshida\\ 
 {\it Argonne National Laboratory, Argonne, IL, USA}~$^{p}$
% ...
}
\newpage 
\begin{supertabular}[h]{rp{14cm}} 
$^{\    1}$ & now visiting scientist at DESY \\ 
$^{\    2}$ & also at IROE Florence, Italy \\ 
$^{\    3}$ & now at Univ. of Salerno and INFN Napoli, Italy \\ 
$^{\    4}$ & also at DESY \\ 
$^{\    5}$ & supported by Worldlab, Lausanne, Switzerland \\ 
$^{\    6}$ & now a self-employed consultant \\ 
$^{\    7}$ & now at Spectral Design GmbH, Bremen \\ 
$^{\    8}$ & now at EDS Electronic Data Systems GmbH, Troisdorf, Germany \\ 
$^{\    9}$ & also at University of Hamburg, Alexander von 
Humboldt Research Award\\ 
$^{  10}$ & now at Dongshin University, Naju, Korea \\ 
$^{  11}$ & now at NASA Goddard Space Flight Center, Greenbelt, MD 
20771, USA\\ 
$^{  12}$ & now at Greenway Trading LLC \\ 
$^{  13}$ & supported by the Polish State Committee for 
Scientific Research, grant No. 2P03B14912\\ 
$^{  14}$ & now at Massachusetts Institute of Technology, Cambridge, MA, 
USA\\ 
$^{  15}$ & visitor from Florida State University \\ 
$^{  16}$ & now at Fermilab, Batavia, IL, USA \\ 
$^{  17}$ & now at ATM, Warsaw, Poland \\ 
$^{  18}$ & now at CERN \\ 
$^{  19}$ & now at ESG, Munich \\ 
$^{  20}$ & now an independent researcher in computing \\ 
$^{  21}$ & now at University of Edinburgh, Edinburgh, U.K. \\ 
$^{  22}$ & PPARC Advanced fellow \\ 
$^{  23}$ & visitor of Univ. of Crete, Greece, 
partially supported by DAAD, Bonn - Kz. A/98/16764\\ 
$^{  24}$ & on leave from MSU, supported by the GIF, 
contract I-0444-176.07/95\\ 
$^{  25}$ & supported by DAAD, Bonn - Kz. A/98/12712 \\ 
$^{  26}$ & supported by an EC fellowship \\ 
$^{  27}$ & PPARC Post-doctoral fellow \\ 
$^{  28}$ & now at Osaka Univ., Osaka, Japan \\ 
$^{  29}$ & also at University of Tokyo \\ 
$^{  30}$ & now at Wayne State University, Detroit \\ 
$^{  31}$ & supported by an EC fellowship number ERBFMBICT 972523 \\ 
$^{  32}$ & now at HERA-B/DESY supported by an EC fellowship 
No.ERBFMBICT 982981\\ 
$^{  33}$ & supported by the Comunidad Autonoma de Madrid \\ 
$^{  34}$ & now at debis Systemhaus, Bonn, Germany \\ 
$^{  35}$ & now at Loma Linda University, Loma Linda, CA, USA \\ 
$^{  36}$ & partially supported by the Foundation for German-Russian 
Collaboration DFG-RFBR (grant no. 436 RUS 113/248/3 and 
no. 436 RUS 113/248/2)\\ 
$^{  37}$ & now at University of Florida, Gainesville, FL, USA \\ 
$^{  38}$ & supported by the Feodor Lynen Program of the Alexander 
von Humboldt foundation\\ 
$^{  39}$ & now with Physics World, Dirac House, Bristol, U.K. \\ 
$^{  40}$ & partly supported by Tel Aviv University \\ 
$^{  41}$ & an Alexander von Humboldt Fellow at University of Hamburg \\ 
$^{  42}$ & supported by a MINERVA Fellowship \\ 
$^{  43}$ & now at ICEPP, Univ. of Tokyo, Tokyo, Japan \\ 
$^{  44}$ & present address: Tokyo Metropolitan University of 
Health Sciences, Tokyo 116-8551, Japan\\ 
$^{  45}$ & now also at Universit\`a del Piemonte Orientale, I-28100 Novara, 
Italy, and Alexander von Humboldt fellow at the University of Hamburg\\ 
$^{  46}$ & supported by the Polish State 
Committee for Scientific Research, grant No. 2P03B09308\\ 
$^{  47}$ & now at University of Rochester, Rochester, NY, USA \\ 
\end{supertabular}
                                                           %
                                                           %
\newpage   
                                                           %
                                                           %
\begin{tabular}[h]{rp{14cm}} 
$^{a}$ &  supported by the Natural Sciences and Engineering Research 
          Council of Canada (NSERC)  \\ 
$^{b}$ &  supported by the FCAR of Qu\'ebec, Canada  \\ 
$^{c}$ &  supported by the German Federal Ministry for Education and 
          Science, Research and Technology (BMBF), under contract 
          numbers 057BN19P, 057FR19P, 057HH19P, 057HH29P, 057SI75I \\ 
$^{d}$ &  supported by the MINERVA Gesellschaft f\"ur Forschung GmbH, the 
German Israeli Foundation, and by the Israel Ministry of Science \\ 
$^{e}$ &  supported by the German-Israeli Foundation, the Israel Science 
          Foundation, the U.S.-Israel Binational Science Foundation, and by 
          the Israel Ministry of Science \\ 
$^{f}$ &  supported by the Italian National Institute for Nuclear Physics 
          (INFN) \\ 
$^{g}$ &  supported by the Japanese Ministry of Education, Science and 
          Culture (the Monbusho) and its grants for Scientific Research \\ 
$^{h}$ &  supported by the Korean Ministry of Education and Korea Science 
          and Engineering Foundation  \\ 
$^{i}$ &  supported by the Netherlands Foundation for Research on 
          Matter (FOM) \\ 
$^{j}$ &  supported by the Polish State Committee for Scientific Research, 
          grant No. 115/E-343/SPUB/P03/154/98, 2P03B03216, 2P03B04616, 
          2P03B10412, 2P03B05315, 2P03B03517, and by the German Federal Ministry
          of Education and Science, Research and Technology (BMBF) \\ 
$^{k}$ &  supported by the Polish State Committee for Scientific 
          Research (grant No. 2P03B08614 and 2P03B06116) \\ 
$^{l}$ &  partially supported by the German Federal Ministry for 
          Education and Science, Research and Technology (BMBF)  \\ 
$^{m}$ &  supported by the Fund for Fundamental Research of Russian Ministry 
          for Science and Edu\-cation and by the German Federal Ministry for 
          Education and Science, Research and Technology (BMBF) \\ 
$^{n}$ &  supported by the Spanish Ministry of Education 
          and Science through funds provided by CICYT \\ 
$^{o}$ &  supported by the Particle Physics and 
          Astronomy Research Council \\ 
$^{p}$ &  supported by the US Department of Energy \\ 
$^{q}$ &  supported by the US National Science Foundation \\ 
\end{tabular} 
} 

\newpage
\pagenumbering{arabic} 

                                                           %

\pagestyle{plain}
\section{Introduction}
\label{sec-int}

The HERA $ep$ collider has extended the kinematic range available for the study
of deep inelastic scattering (DIS) by two orders of magnitude to values of
$Q^2$ up to about $50000\gev^2$, where $Q^2$ is the negative square of the
four--momentum transfer between the lepton and proton. Measurements in this
domain allow new searches for physics processes beyond the Standard Model (SM)
at characteristic mass scales in the $\tev$ range.  A wide class of such
hypothesized new interactions would modify the differential DIS cross sections
in a way which can be parameterized by effective four--fermion contact
interactions (CI) which couple electrons to quarks. This analysis was stimulated
in part by an excess of events over the SM expectation for $Q^2\gtrsim20000
\gev^2$ reported recently by the ZEUS \cite{zeus-97-03} and H1 \cite{h1-97-06}
collaborations, for which CI scenarios have been suggested as possible
explanations (see e.g.\ \citeCIphe).

In a recent publication \cite{zeus-nc-pap}, the Born cross sections at $Q^2>
400\gev^2$ extracted from $47.7\pbi$ of ZEUS neutral--current (NC) $e^+p$ DIS
data collected during the years 1994--1997 have been compared to the SM
predictions primarily derived from measurements at lower $Q^2$, extrapolated to
the HERA kinematic regime. The agreement is generally good. The only discrepancy
is due to the high--$Q^2$ excess of events in the data taken before 1997, which
has not been corroborated by the 1997 data but is still present in the combined
sample with reduced significance.  The present paper presents a comparison of
the full data sample to SM predictions modified by various hypothetical $eeqq$
CI scenarios. Limits on the CI strength and on the effective mass scale
$\Lambda$ are determined for the different CI types.

Limits on $eeqq$ CI parameters have been reported previously by the H1
collaboration \cite{h1-95-09}, by the LEP collaborations (ALEPH
\cite{bar-99-01}, DELPHI \cite{abr-99-01}, L3 \citeCILthree, OPAL
\cite{abb-98-01}), and by the Tevatron experiments CDF \cite{abe-97-03} and
\DO~\cite{abb-98-02}.

This paper is organized as follows: after a synopsis of the relevant theoretical
aspects in \sec{the}, the experimental setup, the event selection and
reconstruction procedures, and the Monte Carlo (MC) simulation are discussed in
\sec{exp}. The CI analysis methods are presented in \sec{ana}, and the results
are summarized in \sec{res}. A discussion of the statistical issues related to
the limit setting and the information needed to combine the results of this
analysis with those of other experiments are given in the Appendix.

\section{Contact--Interaction Scenarios}
\label{sec-the}

\begin{figure}[p]
\vfill
\centerline{\epsfxsize=10cm \epsffile{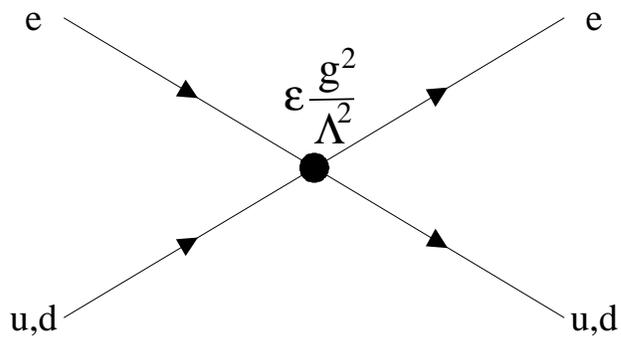}}
\caption{
The Feynman diagram for an $eeqq$ contact interaction.}
\label{fig-contfey}
\vfill
\end{figure}

A broad range of hypothesized non--SM processes at mass scales beyond the HERA
center--of--mass energy, $\sqrt s=300\gev$, can be approximated in their
low--energy limit by $eeqq$ contact interactions (\fig{contfey}), analogous to
the effective four--fermion interaction describing the weak force at low
energies \cite{fer-34-01,*fer-34-02}.  Examples include the exchange of heavy
objects with mass $M\gg\sqrt{s}$ such as leptoquarks or vector bosons
\cite{hab-91-01} and the exchange of common constituents between the lepton and
quark in compositeness models \cite{eic-83-01,ruc-83-01}. Note that the CI
approach is an effective theory which is not renormalizable and is only
asymptotically valid in the low--energy limit.

In the presence of $eeqq$ CIs which couple to a specific quark flavor ($q$), the
SM Lagrangian ${\cal L}_\SM$ receives the following additional terms
\cite{eic-83-01,ruc-83-01,ruc-84-01,hab-91-01}:
\begin{equation}
\renewcommand{\arraystretch}{1.5}
\setlength{\arraycolsep}{0.5mm}
\begin{array}{lrlrrr}
{\cal L} = {\cal L}_\SM +\epsilon 
  &{\displaystyle g^2\over\displaystyle\Lambda^2}\left[\stru{2.ex}{1.ex}\right.&
    \eta_s^q(\ebar_Le_R)(\qbar_Lq_R)+
    \eta_{s'}^q(\ebar_Le_R)(\qbar_Rq_L)+\text{h.c.}
    &&&\text{scalar}\\
  &+&
    \eta_{LL}^q(\ebar_L\gamma^\mu e_L)(\qbar_L\gamma_\mu q_L)
   +\eta_{LR}^q(\ebar_L\gamma^\mu e_L)(\qbar_R\gamma_\mu q_R) 
    &&&\\
  &+&
    \eta_{RL}^q(\ebar_R\gamma^\mu e_R)(\qbar_L\gamma_\mu q_L)
   +\eta_{RR}^q(\ebar_R\gamma^\mu e_R)(\qbar_R\gamma_\mu q_R)
    &&&\text{vector}\\
  &+& 
    \eta_T^q(\ebar_L\sigma^{\mu\nu}e_R)(\qbar_L\sigma_{\mu\nu}q_R)+\text{h.c.}
    \left.\stru{2.ex}{1.ex}\right]\;,
    &&&\text{tensor}
\end{array}
\renewcommand{\arraystretch}{1.0}
                                                        \label{eq-cilag}
\end{equation}
where the subscripts $L$ and $R$ denote the left-- and right--handed helicity
projections of the fermion fields, $g$ is the overall coupling, and $\Lambda$ is
the effective mass scale.  Since $g$ and $\Lambda$ always enter in the
combination $g^2/\Lambda^2$, we adopt the convention $g^2=4\pi$ so that CI
strengths are determined by the effective mass scale, $\Lambda$.  The overall
sign of the CI Lagrangian is denoted by $\epsilon$ in \eq{cilag}. Note that
$\epsilon=+1$ and $\epsilon=-1$ represent separate CI scenarios which are
related to different underlying physics processes.  The $\eta$ coefficients
determine the relative size and sign of the individual terms. Only the vector
terms are considered in this study since strong limits beyond the HERA
sensitivity have already been placed on the scalar and tensor terms (see
\cite{hab-91-01,alt-97-01,bar-98-01} and references therein). In the following,
only CI scenarios will be discussed for which each of the $\eta_{mn}^q$
($m,n=L,R$) is either zero or $\pm1$.

The relevant kinematic variables for this analysis are $Q^2$, $x$ and $y$,
which are defined in the usual way in terms of the four--momenta of the incoming
positron ($k$), the incoming proton ($P$), and the scattered positron ($k'$) as
$Q^2=-q^2=-(k-k')^2$, $x=Q^2/(2q\cdot P)$, and $y=(q\cdot P)/(k\cdot P)$.  The
leading--order, neutral--current $ep$ SM cross section is given by
\begin{eqnarray}
  {d^2\sigma^\NC(e^\pm p)\over dx\,dQ^2}(x,Q^2)&=&
        {2\pi\alpha^2\over xQ^4}\,\left[\left(1+(1-y)^2\right)\,F_2^\NC\mp
        \left(1-(1-y)^2\right)\,xF_3^\NC\strut\right]\;,
                                                        \label{eq-lonc}\\
  F_2^\NC(x,Q^2)&=&\sum_{q=d,u,s,c,b} A_q(Q^2)\cdot
        \left[xq(x,Q^2)+x\qbar(x,Q^2)\right]\;,           \label{eq-f2nc}\\
  xF_3^\NC(x,Q^2)&=&\sum_{q=d,u,s,c,b} B_q(Q^2)\cdot
        \left[xq(x,Q^2)-x\qbar(x,Q^2)\right]\;,         \label{eq-f3nc}
\end{eqnarray}
where $q(x,Q^2)$ and $\qbar(x,Q^2)$ are the parton distribution functions for
quarks and antiquarks, $\alpha$ is the fine structure constant, and $A_q$ and
$B_q$ are defined as follows:
\begin{equation}
  \begin{split}
  A_q(Q^2)&={1\over2}\left[
        (V^\L_q)^2+(V^\R_q)^2+(A^\L_q)^2+(A^\R_q)^2\right]\;,\\
  B_q(Q^2)&=\phantom{{1\over2}\,}\left[
        (V^\L_q)(A^\L_q)-(V^\R_q)(A^\R_q)\right]\;.\\
  \end{split}
                                                        \label{eq-fcoef}
\end{equation}
Terms resulting from CIs can be included in \eqsto{lonc}{fcoef} by replacing the
SM coefficient functions $V^{L,R}_q$ and $A^{L,R}_q$ with
\begin{equation}
  \begin{split}
        V^m_q&=Q_q-(v_e\pm a_e)\,v_q\,\chi_Z+
              {\epsilon\,Q^2\over2\alpha\Lambda^2}(\eta_{mL}^q+\eta_{mR}^q)\;,\\
        A^m_q&=\phantom{Q_q}-(v_e\pm a_e)\,a_q\,\chi_Z+
              {\epsilon\,Q^2\over2\alpha\Lambda^2}(\eta_{mL}^q-\eta_{mR}^q)\;,\\
        v_f  &=T^\sthr_f-2\stws\,Q_f\;,\\
        a_f  &=T^\sthr_f\;,\\
        \chi_Z  &={1\over4\stws\ctws}{Q^2\over Q^2+M_Z^2}\;.
  \end{split}
                                                        \label{eq-fcoci}
\end{equation}
In \eq{fcoci}, the subscript $m$ is $L$ or $R$; the plus (minus) sign in the
definitions of $V^m_q$ and $A^m_q$ is for $m=L$ ($m=R$). The coefficients $v_f$
and $a_f$ are the SM vector and axial--vector coupling constants of an electron
($f=e$) or quark ($f=q$); $Q_f$ and $T^\sthr_f$ denote the fermion's charge and
third component of the weak isospin; $M_Z$ and $\tw$ are the $Z$ mass and the
Weinberg angle. In the limit $\Lambda\to\infty$, the coefficient functions
$V^m_q$ and $A^m_q$ in \eq{fcoci} reduce to their SM forms.

As can be seen from \eqsto{lonc}{fcoci}, the effect of a CI on the NC DIS
cross section depends on the specific scenario.  In general, two kinds of
additional terms are produced. One kind is proportional to $1/\Lambda^4$ and
enhances the cross section at high $Q^2$. The second is proportional to
$1/\Lambda^2$ and is caused by interference with the SM amplitude, which can
either enhance or suppress the cross section at intermediate $Q^2$. The
predicted ratio (SM+CI)/SM of $d^2\sigma/(dQ^2\,dx)$ depends on $Q^2$ at fixed
$x$, but also on $x$ at fixed $Q^2$ due to the different $y$ dependences of the
coefficient functions multiplying the $F_2$ and $F_3$ terms in \eq{lonc}. Note
that CIs induce modifications of the SM cross section for all $x$ and $Q^2$, in
contrast e.g.\ to the direct production of an $eq$ resonance in the 
$s$--channel.

For $ep$ scattering at HERA, the contribution of second-- and third--generation
quarks to CI cross--section modifications is suppressed by the respective parton
distribution functions in the proton.  For the present analysis, flavor
symmetry,
\begin{equation}
        \eta_{mn}^d=\eta_{mn}^s=\eta_{mn}^b
        \qquad\text{and}\qquad
        \eta_{mn}^u=\eta_{mn}^c
                                                        \label{eq-flasy}
\end{equation}
is assumed unless explicitly stated otherwise. The CI limits reported here are
only weakly sensitive to this assumption.\footnote{A similar statement is true
for the Tevatron CI limits from lepton--pair production, which also depend on
the parton distributions in the proton. In contrast, the CI analyses at LEP are
sensitive to the cross section $\sigma(e^+e^-\to \text{hadrons})$ and the
resulting limits depend strongly on flavor symmetry assumptions.}  Contributions
from the top quark content of the proton are almost completely suppressed due to
the large top mass and are neglected in this analysis.

Using the relations in \eq{flasy}, there are eight independent vector terms in
\eq{cilag}, which lead to a large list of possible CI scenarios. To reduce this
list, we consider the following:
\begin{tightitemize}{1.pt plus 1.pt minus 0.5pt}{0.pt plus 0.5pt minus 1.pt}
\item
Recent measurements of parity--violating transition amplitudes in cesium atoms
\cite{woo-97-01} imply very restrictive constraints on CIs \citeAPVlim. These
limits are avoided by parity--conserving CI scenarios, i.e.\ if
\begin{equation}
  \eta^q_{LL}+\eta^q_{LR}-\eta^q_{RL}-\eta^q_{RR} = 0\;.
                                                        \label{eq-apv}
\end{equation}
Conforming to this constraint, in particular, excludes CIs of purely chiral
type, i.e.\ those for which the $\eta_{mn}^q$ are non--zero only for one
combination of $m$ and $n$.
\item 
$SU(2)_L$ invariance requires $\eta^u_{RL}=\eta^d_{RL}$ \cite{bar-98-01}. Terms
violating this relation are considered only for $u$ quarks, which dominate the
high--$x$ cross section at HERA, and hence show the largest CI--SM interference
effects for a given $\Lambda$.  A CI signal from this source could therefore
manifest itself in $ep$ collisions while avoiding strong $SU(2)$--breaking 
effects e.g.\ at LEP.
\end{tightitemize}

Based on these considerations, the 30~specific CI scenarios listed in
\tab{models} are explored in this paper. Note that each line in this table
represents two scenarios, one for $\epsilon=+1$ and one for $\epsilon=-1$
(denoted as VV$^+$, VV$^-$ etc.). All scenarios respect
\eq{apv}, and all scenarios except U2, U4 and U6 obey $SU(2)$ symmetry. The
$SU(2)$--conserving CI scenarios with $\eta^u_{LL}\ne\eta^d_{LL}$ (U1 and U3)
would also induce an $e\nu qq$ CI signal in charged--current (CC) DIS,
$e^+p\to\nubar X$.  We have not used the CC data sample to constrain further
these scenarios.  

\begin{figure} [p]
  \centerline{\psfig{file=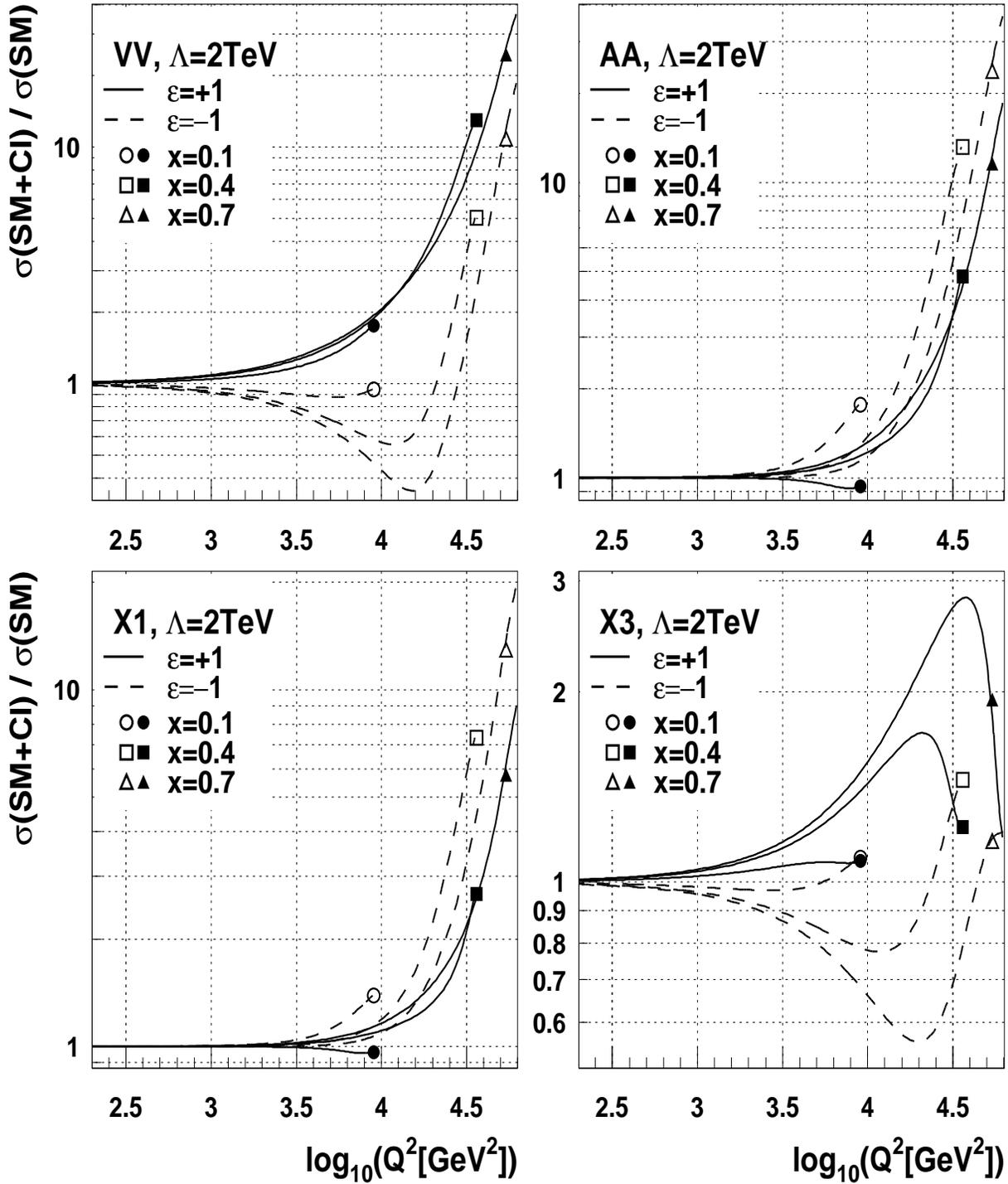,width=\textwidth,height=19.cm}}
  \vspace{-2.mm}
  \caption{Examples of the relative influence of a CI on the NC DIS
           cross section $d^2\sigma/(dx\,dQ^2)$ at various fixed values of
           $x$. The ratio (SM+CI)/SM of differential cross sections for the
           scenarios VV (top left), AA (top right), X1 (bottom left) and X3
           (bottom right) is plotted for $\Lambda=2\tev$. The symbols are used
           only to label the different curves.}
  \label{fig-cixsect}
\end{figure}

Several examples of modifications of the SM cross sections by CIs are
illustrated in \fig{cixsect}.  The cross--section modifications for the X1--X6
and the corresponding U1--U6 scenarios are similar, demonstrating that the $d$
quarks have little impact on the CI analysis.

\section{Experimental Setup and Data Samples}
\label{sec-exp}

This analysis uses the data samples, Monte Carlo simulation, event selection,
kinematic reconstruction, and assessment of systematic effects used in the NC
DIS analysis described in \cite{zeus-nc-pap}. The data were collected during the
years 1994--1997 in $e^+p$ collisions with beam energies $E_e=27.5\gev$ and
$E_p=820\gev$. The relevant aspects of the experimental setup, event selection,
and reconstruction are summarized briefly below. More details can be found in
\cite{zeus-nc-pap}.

The ZEUS detector is described in detail elsewhere \cite{zeus-bb}. The main
components used in the present analysis are the central tracking detector (CTD)
\citeCTD, positioned in a $1.43\,$T solenoidal magnetic field, and the
compensating uranium--scintillator sampling calorimeter (CAL) \citeCAL,
subdivided into forward (FCAL), barrel (BCAL) and rear (RCAL) sections.  Under
test beam conditions, the CAL energy resolution is $\sigma(E)/E=18\%/\sqrt{E\,
[\Gev]}$ for electrons and $35\%/\sqrt{E\,[\Gev]}$ for hadrons. A three--level
trigger is used to select events online. The trigger decision is based mainly on
energies deposited in the calorimeter, specifically on the electromagnetic
energy, on the total transverse energy, and on\footnote{ZEUS uses a
right--handed Cartesian coordinate system centered at the nominal interaction
point, with the $Z$ axis pointing in the proton beam direction. The polar angle
is defined with respect to this system in the usual way.}  \hbox{$E-p_Z=\sum_i
E_i(1-\cos\theta_i)$} (the sum running over all calorimeter energy deposits).
For fully contained events, the expected value of $E-p_Z$ is given by
$2E_e=55\gev$.  Timing cuts are used to reject beam--gas interactions and cosmic
rays.

The luminosity is measured to a precision of $1.6\%$ from the rate of energetic
brems\-strah\-lung photons produced in the process $ep\to ep\gamma$
\cite{zeus-det-92-12,*zeus-94-02}.

The offline event reconstruction applies an algorithm to identify the scattered
positron using the topology of its calorimeter signal and the tracking
information. The measured energies are corrected for energy loss in inactive
material between the interaction point and the calorimeter, for calorimeter
inhomogeneities, and for effects caused by redirected hadronic energy from
interactions in material between the primary vertex and the calorimeter or by
backsplash from the calorimeter (albedo).  The kinematic variables for NC DIS
candidate events are calculated from the scattering angle of the positron and
from an angle representing the direction of the scattered quark. The latter is
determined from the transverse momentum and the $E-p_Z$ of all energy deposits
except those assigned to the scattered positron.

The appropriately corrected experimental quantities are used to make the offline
event selection.  The major criteria are \cite{zeus-nc-pap}: (i) the event
vertex must be reconstructed from the tracking information, with $|Z_{\rm
vtx}|<50\cm$; (ii) an isolated scattered positron with energy $E_e'>10\gev$
has to be identified; (iii) $38\gev<E-p_Z<65\gev$; (iv) $y_e<0.95$, where $y_e$
is the value of $y$ as reconstructed from the measured energy and angle of the
scattered $e^+$. The requirements (iii) and (iv) reject background events from
photoproduction.

Monte Carlo simulations are used to model the expected distributions of the
kinematic variables $x$, $y$ and $Q^2$ and to estimate the rate of
photoproduction background events. NC DIS events including radiative effects are
simulated using the {\sc heracles~4.5.2} \cite{kwi-92-01,*spi-96-01} program
with the {\sc django~6.24} \cite{cha-94-01,*spi-96-02} interface to the
hadronization programs.  In {\sc heracles}, corrections for initial-- and
final--state radiation, vertex and propagator corrections, and two--boson
exchange are included.  The underlying cross sections are calculated in
next--to--leading order QCD using the CTEQ4D\footnote{The final versions of the
CTEQ5 \cite{lai-99-01} and MRST \cite{mar-98-01,*mar-99-01} PDF sets became
available only after completion of this analysis.} set \cite{lai-97-02} of
parton distribution functions (PDFs).  The NC DIS hadronic final state is
simulated using the color--dipole model of {\sc ariadne~4.08} \cite{lon-92-01}
and, as a systematic check, the {\sc meps} option of {\sc lepto~6.5}
\cite{ing-97-01} for the QCD cascade. Both programs use the Lund string model of
{\sc jetset~7.4} \citeJETSET for the hadronization. MC samples of
photoproduction background events are produced using the {\sc herwig~5.8}
\cite{mar-92-01} generator.  All MC signal and background events are passed
through the detector simulation based on {\sc geant} \cite{bru-87-01},
incorporating the effects of the trigger. They are subsequently processed with
the same reconstruction and analysis programs used for the data. All MC events
are weighted to represent the same integrated luminosity as the experimental
data.

Good agreement is found in \cite{zeus-nc-pap} both between the distributions of
kinematic variables in data and MC, and between the measured differential
cross sections $d\sigma/dQ^2$, $d\sigma/dx$ and $d\sigma/dy$ and the respective
SM predictions, with the possible exception of the two events at $Q^2>35000
\gev^2$.

\textheight     660.pt
\section{Analysis Method}
\label{sec-ana}

The CI analysis compares the distributions of the measured kinematic variables
with the corresponding distributions from a MC simulation of events of the type
$e^+p\to e^+X$, with the weight
\begin{equation}
  w=\left.{{d^2\sigma\over dx\,dQ^2}(\text{SM$+$CI})\over
          {d^2\sigma\over dx\,dQ^2}(\text{SM})}\right|_{{\rm true\ } x,y,Q^2}
                                                        \label{eq-wdef}
\end{equation}
applied to each reconstructed MC event in order to simulate the CI scenarios.
The weight $w$ is calculated as the ratio of leading--order\footnote{Note that
CIs are a non--renormalizable effective theory for which higher orders are not
well--defined. Radiative corrections due to real photon emission are expected to
cancel to a large extent in \eq{wdef}.} cross sections, evaluated at the
``true'' values of $x$, $y$ and $Q^2$ as determined from the four--momentum of
the exchanged boson and the beam momenta. In cases where a photon with energy
$E_\gamma$ is radiated off the incoming positron (initial--state radiation), the
$e$ beam energy is reduced by the energy of the radiated photon.  The
reweighting procedure using \eq{wdef} accounts correctly for correlations
between the effects of a CI signal and the pattern of acceptance losses and
migrations.

The simulated background events from photoproduction are added to the selected
NC--DIS MC data sets. The photoproduction contamination is highest at high $y$
and is estimated to be less than $0.5\%$ overall and below $3\%$ in any of the
bins used for the cross--section measurements in \cite{zeus-nc-pap}.

For each of the CI scenarios, two statistical methods are used. Each
incorporates a log--likelihood function\footnote{A discussion of a probabilistic
interpretation of the log--likelihood function based on a Bayesian approach can
be found in the Appendix.}
\begin{equation}
  L(\epslam)=-\sum_{i\in\text{data}}\log{p_i(\epslam)}\;,
                                                        \label{eq-lldef}
\end{equation}
where the $p_i$ are appropriately normalized probabilities which are derived
from a comparison of measured and simulated event distributions ($i$ runs over
individual events in method~1 and over bins of a histogram in method~2, see
below).  Note that the two CI scenarios corresponding to two sets of
$\eta_{mn}^q$ values differing only in the overall sign $\epsilon$ are combined
into one log--likelihood function.  A description of the data samples used for
evaluating $L(\epslam)$ is given in \tab{dasamp} for both methods.

\begin{figure}[p]
  \centerline{\psfig{file=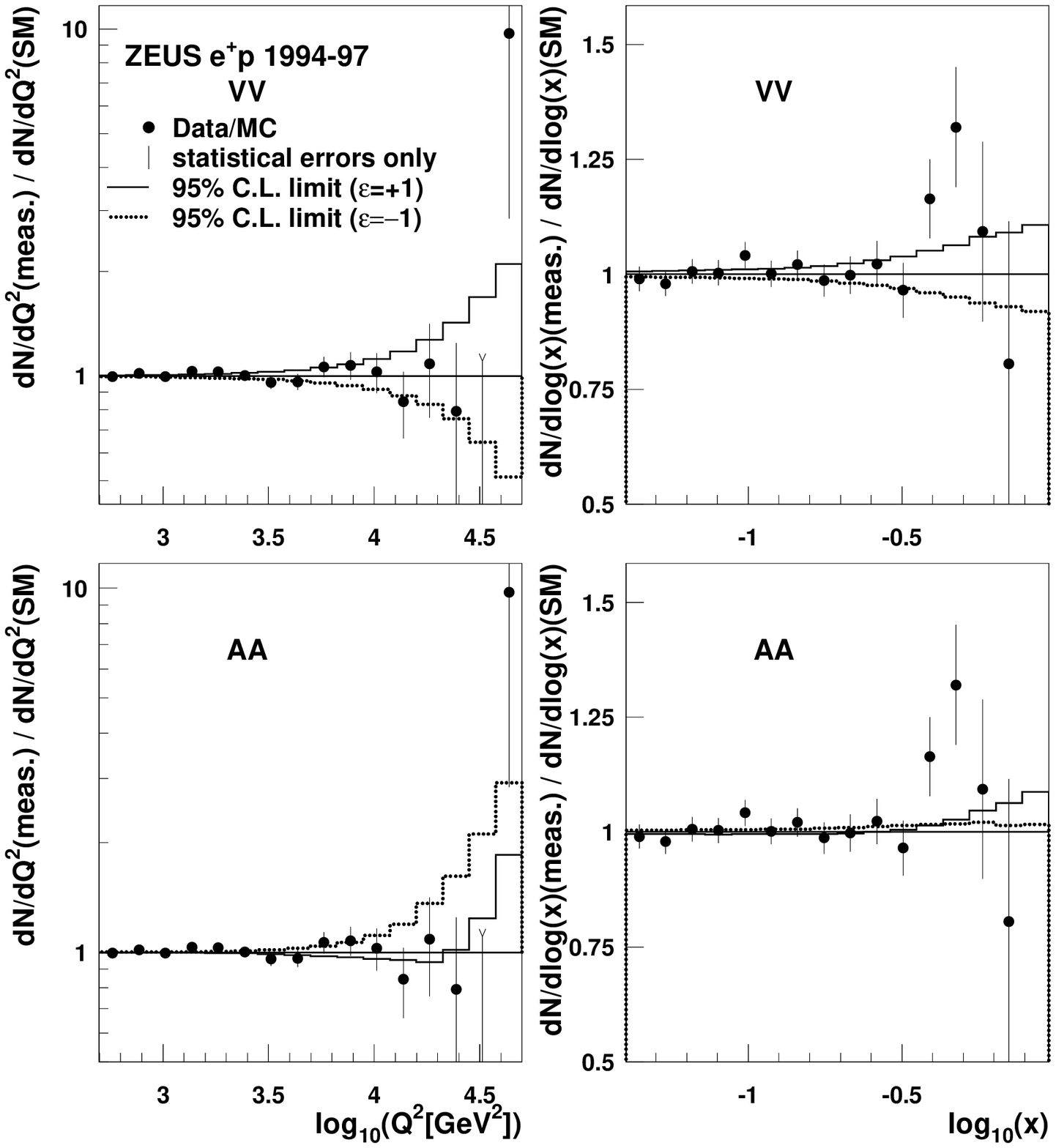,width=\textwidth,height=19.cm}}
  \vspace{-2.mm}
  \caption{Exclusion limits on CI strengths in terms of the corresponding
           modification of the expected distributions of $Q^2$ (left) and $x$
           (right) in the kinematic region used for method~1 (see \sec{ana}
           and \tab{dasamp}).  The dots represent the ratios of observed and
           expected numbers of events and the error bars indicate the
           statistical uncertainties. The histograms show the modification of
           this ratio for the VV (top) and AA (bottom) scenarios as obtained
           with the $\Lambda_{\rm lim}$ values given in \tab{reszeus}. The
           solid (dotted) line corresponds to the $95\%$ exclusion limit for
           $\epsilon=+1$ ($\epsilon=-1$).}
  \label{fig-xqdisa}
\end{figure}

\textheight     670.pt
\begin{itemize}
\item
\uline{Unbinned fit to the $(x,y)$--distribution:}

The available experimental information entering the analysis can be split into
two parts, the shape of the $(x,y)$ distribution, $(d^2N/dx\,dy)/N_\tot$, and
the total number of events, $N_\tot$. The latter is related to the total cross
section $\sigma_\tot(\epslam)$ in the kinematic region under study by
$N_\tot=\epsilon_\tot\cdot\lum\cdot\sigma_\tot$, where $\epsilon_\tot$ is the
average acceptance and $\lum$ denotes the integrated luminosity.

In the first method, an unbinned log--likelihood technique is applied to
calculate $L_\sone(\epslam)$ from the individual kinematic event coordinates
$(x_i,y_i)$. This method only makes use of the shape of the $(x,y)$
distribution.  The sum in \eq{lldef} runs over all events in the selected data
sample.  The MC events are appropriately reweighted to simulate a CI scenario
with strength $\epslam$, as outlined in~\eq{wdef}.  The probability density
$p(x,y;\epslam)$ required to calculate $p_i(\epslam)=p(x_i,y_i;\epslam)$ is
determined from the resulting density of MC events in $x$ and $y$ and is
normalized to unity, thereby discarding the information on $\lum\cdot\sigma_\tot
(\epslam)$. The justification for this deliberate reduction of experimental
information is given {\it a posteriori} by the fact that $\sigma_\tot(\epslam)$
depends only weakly on $\Lambda$ in the parameter space of interest: for
$\Lambda$ values larger than the $95\%$ lower exclusion limits (see \sec{res}),
$\sigma_\tot(\epslam)$ deviates from the SM value, $\sigma_\tot(0)$, by less
than $2\%$ for all scenarios except X1 and X6, for which $2.2\%$ and $2.6\%$ are
reached, respectively. This sensitivity is smaller than, or of the same order
as, the $1.6\%$ systematic luminosity uncertainty quoted in \sec{exp} and is
hence not significant.

Even though $N_\tot$ is not used in this method, it is important to note that
data ($N_\tot=13243$ events observed) and SM prediction ($13151$ events
expected) agree within the luminosity uncertainty and the statistical
error. Furthermore, \figand{xqdisa}{xqdisb} demonstrate that the $Q^2$ and $x$
distributions agree in shape with the SM expectation. Therefore, a deterioration
of the agreement of data and expectation with increasing $|\epslam|$, indicated
by an increase in the log-likelihood function with respect to a minimum close to
$\epslam=0$, can be interpreted in terms of CI exclusion limits on $\epslam$
or~$\Lambda$.

The sensitivity of the results to systematic effects is studied by repeating the
limit setting procedure (see below) for analysis parameters and selection
requirements which are varied within admissible ranges.  These systematic checks
include those which were performed in the underlying cross--section analysis
\cite{zeus-nc-pap}:
\begin{tightsubitemize}{0.pt plus 1.pt minus 0.5pt}{0.pt plus 0.5pt minus 3.pt}
\item
use of MC samples generated with the {\sc meps} instead of the {\sc ariadne}
option (as described in \sec{exp});
\item
variations of trigger or reconstruction efficiencies and of experimental
resolutions within their uncertainties by suitably reweighting the MC events;
\item
modifications of the cuts and parameters used for event reconstruction and
selection;
\item
variation of the calorimeter energy scales in the analysis of the data but not
in that of the MC events.
\end{tightsubitemize}
In addition, systematic uncertainties related to the CI fitting procedure are
investigated by
\begin{tightsubitemize}{0.pt plus 1.pt minus 0.5pt}{0.pt plus 0.5pt minus 3.pt}
\item
use of MC samples which were generated with the following alternatives to the
CTEQ4D PDF set: (i) with the PDF set MRSA \cite{mar-95-01} and (ii) using the
results of a recent NLO QCD fit \cite{bot-99-01} to the 1994 ZEUS
\cite{zeus-96-07} and H1 \cite{h1-96-07} structure function data and to
fixed--target data;
\item
use of MC samples generated with the PDF sets CTEQ4A2 and CTEQ4A4
\cite{lai-97-02}, corresponding to values of $\als(M_Z^2)=0.113$ and $0.119$
(instead of $\als(M_Z^2)=0.116$ used for CTEQ4D);
\item
changing the amount of photoproduction background in the MC sample by
$\pm100\%$;
\item
modifying details of the method used to infer $p(x,y;\epslam)$ from
the MC event distributions;
\item
calculating the CI cross sections in the following alternative ways:
\begin{tightssubitemize}{0.pt plus 1.pt minus 0.5pt}{0.pt plus 0.5pt minus 3.pt}
\item
with different parton distribution functions,
\item
using NLO instead of LO QCD calculations and parton distributions,
\item
with the couplings restricted to first--generation quarks ($\eta_{mn}^s
\eql\eta_{mn}^b\eql\eta_{mn}^c=0$),
\item
with the couplings restricted to first-- and second--generation quarks
($\eta_{mn}^b\eql0$).
\end{tightssubitemize}
\end{tightsubitemize}

The procedure to determine limits on the CI parameters including the information
from the systematic checks is discussed at the end of this section.

\item
\uline{Binned fit to the $Q^2$ distribution:}

In the second method, $L_\stwo(\epslam)$ is determined from the $Q^2$
distributions, using Poisson statistics for the numbers of events in each $Q^2$
interval. Here, the sum in \eq{lldef} runs over all $Q^2$ bins.  For the
calculation of $L_\stwo$, both the shape and the normalization of $dN/dQ^2$ are
used.

The systematic uncertainties are included in $L_\stwo$ using the assumptions
that they are fully correlated between bins and that the probability densities
for all uncertainties have Gaussian shapes. The effects taken into account are
equivalent to those described above for the unbinned method and include in
addition a $1.6\%$ uncertainty on the integrated luminosity.
\end{itemize}

\begin{figure}[p]
  \centerline{\psfig{file=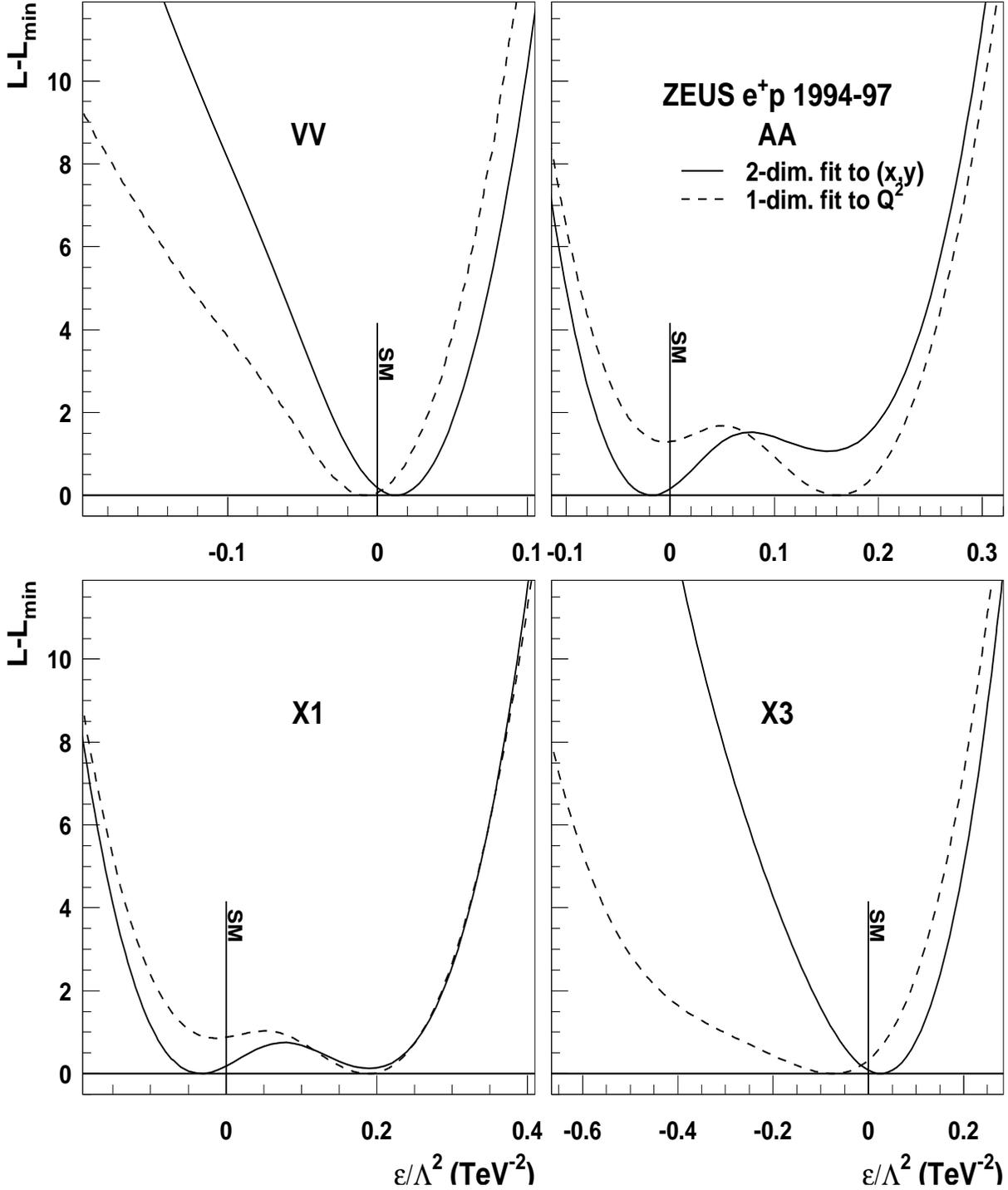,width=\textwidth,height=19.cm}}
  \vspace{-2.mm}
  \caption{Log--likelihood as a function of $\epsilon/\Lambda^2$ as obtained
           by the two methods described in \sec{ana}, for the scenarios VV (top
           left), AA (top right), X1 (bottom left), and X3 (bottom right). For
           both methods, the values of $L-L_\smin$ are shown. Note that the two
           methods use different statistical approaches and, in addition, differ
           in their treatment of systematic effects. Each figure represents two
           CI scenarios ($\epsilon=+1$, $\epsilon=-1$).}
  \label{fig-llcomp}
\end{figure}

Both methods have been shown to provide unbiased estimates of the CI strength
when applied to MC samples.  A few examples of comparisons of the resulting
log--likelihood functions $L_\sone$ and $L_\stwo$ are shown in \fig{llcomp}.
Both functions agree with each other for most of the CI scenarios under study.
However, for a few scenarios such as X2$^-$ and X3$^-$, $L_\sone$ rises faster
than $L_\stwo$ with decreasing $\Lambda$. This can be understood as a
consequence of the fact that $L_\sone$ uses the full two--dimensional
information of the $(x,y)$ distribution and is hence more sensitive to those
scenarios which imply a marked $x$--dependence of the modification to the
cross--section at fixed $Q^2$. The two--minimum structure of the log--likelihood
functions seen in the AA and X1 scenarios in \fig{llcomp} is characteristic for
several CI scenarios for which the destructive SM$\times$CI interference term
cancels approximately the pure CI$\times$CI term in a range of typically
$0.1\tev^{-2}\lesssim|\epslam|\lesssim0.2\tev^{-2}$. MC studies indicate that
the exact shape of $L_{\sone,\stwo}$ in the vicinity of these double--minima is
dominated by the random pattern of statistical fluctuations of the event
distributions, but that the two--dimensional method has a higher probability
than the one-dimensional method to assign a larger value of $L$ to the ``non--SM
minimum'' than to the ``SM--minimum'' (the AA case in \fig{llcomp} is typical).
This is again understood as a consequence of the additional input information
for the two--dimensional method.  The normalization information, $N_\tot/
(\lum\cdot\sigma_\tot(\epslam))$, cannot distinguish between the minima since
the difference of $\sigma_\tot(\epslam)$ between them is much less than the
uncertainty of $\lum$.

\begin{figure}[p]
  \centerline{\psfig{file=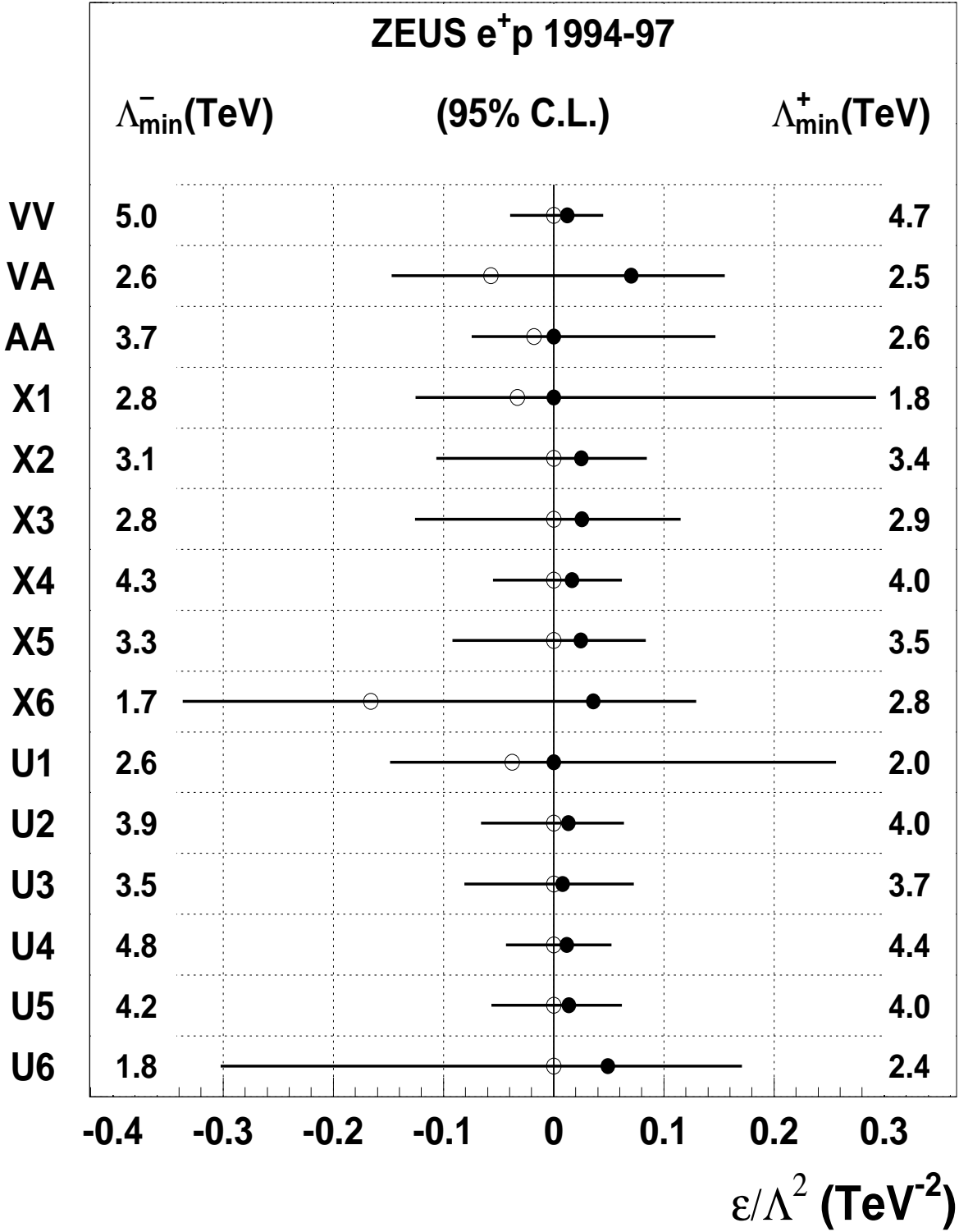,width=15.5cm,height=20.cm}}
  \vspace{-2.mm}
  \caption{Confidence intervals of $\epsilon/\Lambda^2$ at $95\%$ C.L.\ for all
           CI scenarios studied in this paper (horizontal bars). The filled
           (open) circles indicate the positions of the log--likelihood minima,
           $\epsilon/\Lambda_\zero^2$, for $\epsilon=+1$ ($\epsilon=-1$). The
           numbers at the right (left) margin are the lower $\Lambda$ limits for
           $\epsilon=+1$ ($\epsilon=-1$).}
  \label{fig-reszeus}
\end{figure}

The best estimates, $\Lambda_\zero$, for the different CI scenarios are given by
the positions of the respective minima of $L_{\sone,\stwo}(\epslam)$ with
$\epsilon=-1$ and $\epsilon=+1$. The values of $\epsilon/\Lambda_\zero^2$
resulting from the unbinned method are indicated in \fig{reszeus}. Note that
$1/\Lambda_\zero^2=0$ is taken for all cases where $L_{\sone,\stwo}(\epslam)$
rises monotonically for a given scenario, i.e.\ for $\epsilon=+1$ or
$\epsilon=-1$ (see \fig{llcomp}).

An analysis of the log--likelihood functions (referred to as the
``$L$--analysis'' in the following) is usually employed to calculate confidence
level intervals (i.e.\ limits) of $\epslam$. A discussion of some aspects
related to the $L$--analysis approach can be found in the Appendix, where
polynomial parameterizations for $L_\sone(\epslam)$ are also provided, which may
prove useful for combining our results with those from other experiments.  One
problem with the $L$--analysis is that its results depend on the choice of the
``canonical variable'' in terms of which it is performed; for example the
$\Lambda$ limits are different if $L$ is evaluated as a function of
$\epsilon/\Lambda$ instead of $\epslam$. In order to avoid this ambiguity, MC
experiments (MCE) are used, i.e.\ statistically independent MC data samples
corresponding to the data luminosity. The presence of CIs in the MCEs is
simulated by reweighting the events in the MCEs according to \eq{wdef}.  For
each MCE, the log-likelihood analysis is performed as a function of the assumed
``true'' value of $\Lambda$, $\Lambda_\true$, for each of the CI scenarios under
study. The lower limit of $\Lambda$ at $95\%$ C.L.\ for a given CI scenario is
determined as the value of $\Lambda_\true$ at which $95\%$ of the MCEs produce
most likely values of $|\epslam|$ larger than that found in the data.  In the
cases where $1/\Lambda_\zero^2\ne0$, the MCEs are also used to estimate the
probability, $p_\SM$, that a statistical fluctuation in an experiment with the
SM cross section would produce a value of $\Lambda_\zero$ smaller than that
obtained from the data. Note that a high value of $p_\SM$ does not in itself
signify that the SM prediction describes the data well, but indicates that the
inclusion of the CI scenario under study does not significantly improve the
agreement between data and prediction.

For the two--dimensional fitting method, the above procedure is repeated for
each systematic check, using statistically independent MCE sets which reflect
the corresponding modifications of the analysis. Each such MCE set consists of
$500$~MCEs.  The resulting CI limits are scattered around the limits of the
central analysis, with deviations in both directions being about equally
frequent.\footnote{This implies that roughly $50\%$ of all checks produce
limits which are stronger than the central ones and is related to the fact
that, to a very good approximation, $\Lambda$ depends linearly on continuous
parameters, like $\als$, within their uncertainty intervals.} The $\Lambda$
limits deviate from their central values by typically less than $15\%$, though
by as much as $30\%$ in a few cases. The modification of the underlying SM cross
section induced by a variation of $\als(M_Z^2)$ and by using different PDF sets
(see above) causes variations of the $\Lambda$ limits of typically a few percent
and $25\%$ maximally.

Systematic effects are finally taken into account in the CI limit analysis by
combining the MCE sets of all systematic checks and determining the values of
$\Lambda_\true$ for which $95\%$ of all MCEs in the combined set produce most
likely values of $|\epslam|$ larger than that found in the data. This procedure
is an approximation to averaging over the spectra of systematic effects,
assuming that the different checks are uncorrelated and that the ranges of
parameter variations (e.g.\ of the calorimeter energy scales or of $\alsmz$)
reflect the actual uncertainties.  The corresponding question defining a $95\%$
C.L.\ limit is: ``which value of $\Lambda_\true$ causes deviations from the SM
prediction which are larger than that observed in the experimental data in
$95\%$ of all ZEUS--type experiments exhibiting systematic differences according
to the spectra determined in the analysis of systematic effects''.\footnote{This
corresponds to assigning equal {\it a priori} probabilities to each of the
tested variations and reflects the fact that by construction neither of them can
be excluded or favored.}

The $\Lambda$ limits resulting from both log--likelihood methods agree to within
$15\%$ in all cases except for the scenarios AA$^+$, X2$^-$ and X3$^-$, for
which the two--dimensional method has higher sensitivity and correspondingly
yields significantly stronger limits. Therefore, the results of this method are
presented in the following.

\section{Results}
\label{sec-res}

The resulting SM probabilities $p_\SM$ (see \tab{reszeus}) do not indicate
significant amplitudes for any of the CI scenarios considered.  Therefore,
we report upper limits on $1/\Lambda^2$ and the corresponding lower limits on
$\Lambda$.

A selection of plots demonstrating the expected modifications of the $Q^2$-- and
$x$--distri\-butions in the presence of CIs with strengths corresponding to the
$95\%$ C.L.\ exclusion limits is shown in \figand{xqdisa}{xqdisb}.\footnote{Note
that the binning of the data in $x$ and $Q^2$ used for this presentation is
irrelevant in the analysis of the limits provided in \tab{reszeus}.}  As
mentioned in \sec{int}, the data show an excess over the SM predictions at
$Q^2\gtrsim35000\gev^2$. However, it is apparent that CIs cannot provide an
improved description of this excess while simultaneously describing the data
well at lower $Q^2$, where data and SM expectation are in good agreement.  These
figures also confirm that the $x$--dependence of the (SM+CI)/SM cross--section
ratio differs markedly between different CI scenarios and obviously contributes
to the sensitivity of the CI fit, e.g.\ in the X3 case.  This statement is
generally true for all cases where the limits derived from the two analysis
methods differ significantly.

The lower limits on $\Lambda$ ($\Lambda^\pm_\slim$) and the probabilities
$p_\SM$ are summarized in \tab{reszeus} and are displayed in \fig{reszeus}. In
none of the cases does the SM probability fall below $16\%$.  The $\Lambda$
limits range from $1.7\tev$ to $5\tev$. Those few cases with limits below
$2.5\tev$ correspond to log--likelihood functions having a broad minimum, either
in the region with $\epsilon=+1$ (X1, U1) or with $\epsilon=-1$ (X6, U6); these
minima correspond to parameter combinations for which the pure CI$\times$CI
contribution and the CI$\times$SM interference term approximately cancel in the
HERA kinematic regime.\footnote{Note that these cancellations happen at opposite
$\epsilon$ for $e^-p$ scattering.}

\Tab{ciother} shows a comparison of the ZEUS CI results with corresponding
limits reported recently by other experiments which study $eeqq$ CIs in $e^+e^-$
scattering at LEP (ALEPH \cite{bar-99-01}, L3 \citeCILthreecomp, OPAL
\cite{abb-98-01}) or via Drell--Yan pair production in $p\pbar$ scattering (CDF
\cite{abe-97-03}, \DO \cite{abb-98-02}).  The H1 \cite{h1-95-09} and DELPHI
\cite{abr-99-01} collaborations report results only for purely chiral CIs which
cannot be compared to the results of this paper.  All limits shown in
\tab{ciother} have been derived assuming flavor symmetry (see \eq{flasy}),
except the LEP limits for the U3 and U4 scenarios which are for
first--generation quarks only. Limits for the X2, X5, U1, U2, U5 and U6
scenarios are not included in \tab{ciother} because there exist no previously
published results.  ZEUS and the other experiments are all sensitive to CIs at
mass scales of a few $\tev$. The relative sensitivity to different CI scenarios
depends on the CI$\times$SM interference sign which is opposite in $e^+p$
scattering on the one hand and in $e^+e^-$ and $p\pbar$ scattering on the
other.\footnote{The CI limits from CDF and the LEP experiments have been quoted
here according to the sign convention used in the cited papers.} Where
available, the LEP limits often exceed the results of this paper, although it
should be noted that this depends on the assumption of a flavor--symmetric CI
structure. Limits for CIs which couple only to first--generation quarks would
differ only by a small amount from those reported here for $ep$ or $p\pbar$
scattering, but would be significantly weaker in the case of LEP.

\textheight     650.pt
\section{Conclusions}
\label{sec-con}

We have searched for indications of $eeqq$ contact interactions in $47.7\pbi$ of
ZEUS high--$Q^2$ $e^+p$ neutral--current deep inelastic scattering data. The
distributions of the kinematic variables in the data have been compared to
predictions derived for 30~scenarios of vector contact interactions which differ
in their helicity structure and quark--flavor dependence.  In none of the cases
has a significant indication of a contact interaction been found and $95\%$
C.L.\ upper limits on $1/\Lambda^2$ have been determined for each of these
scenarios.  The lower limits on $\Lambda$ range between $1.7\tev$ and $5\tev$
and are found to be largely independent of the statistical method applied.

The results exhibit a sensitivity to contact interactions similar to that
recently reported by other experiments; in order to allow full use to be made of
the available experimental data, the information needed to combine the results
of this analysis with those from other sources is provided.  Some of the limits
reported here are the most restrictive yet published, and several of the
contact--interaction scenarios have been studied in this paper for the first
time.

\vskip5.mm
{\it Acknowledgments:}
We appreciate the contributions to the construction and maintenance of the ZEUS
detector by many people who are not listed as authors.  We especially thank the
DESY computing staff for providing the data analysis environment and the HERA
machine group for their outstanding operation of the collider. Finally, we thank
the DESY directorate for strong support and encouragement.
 
This paper was completed shortly after the tragic and untimely death of
Prof.~Dr.\ B.~H.~Wiik, Chairman of the DESY directorate. All members of the ZEUS
collaboration wish to acknowledge the remarkable r\^ole which he played in the
success of both the HERA project and of the ZEUS experiment. His inspired
scientific leadership, his warm personality and his friendship will be sorely
missed by us all.

\clearpage
\section*{Appendix: The Log--Likelihood Functions}
\label{sec-app}

In this Appendix we summarize some aspects of interpreting the log--likelihood
functions $L(\epslam)$ using a Bayesian probabilistic approach. The results of
the unbinned method described in \sec{ana} have been employed here, but
systematic effects have not been taken into account. For simplicity, we will
denote the log--likelihood functions by $L$ instead of $L_\sone$ in this
Appendix.

For ease of calculation, the functions $L(\epslam)-L_\smin$ have been
parameterized as eighth--order polynomials in the region where $L(\epslam)
-L_\smin<18$, corresponding approximately to a $\pm6\sigma$ interval around the
minimum of $L$. The polynomial coefficients are summarized in \tab{polco}.  The
accuracy of the parameterizations is typically better than $0.1$~units in $L$.
Note that, neglecting systematic effects, these parameterizations allow one to
combine the ZEUS results with those of other experiments by simply adding the
$L$ functions and repeating the analysis described below.

The Bayesian approach starts from the relation
\begin{equation}
  p(\epslam|{\cal D})\propto p({\cal D}|\epslam)\cdot p_\zero(\epslam)\;,
                                                        \label{eq-baypro}
\end{equation}
where ${\cal D}$ symbolizes the experimental data, $p({\cal D}|\epslam)$ is the
conditional probability to observe ${\cal D}$ for a given value of $\epslam$,
and $p_\zero(\epslam)$ is the {\it prior probability} describing the knowledge
about $\epslam$ before the experiment was conducted. The probability
$p(\epslam|{\cal D})$ assigned to $\epslam$ under the condition of having
observed ${\cal D}$ is what we actually want to derive. 

In the following, we will identify
\begin{equation}
  p({\cal D}|\epslam)\propto\exp\left(-L(\epslam)\right)\;,
                                                        \label{eq-baydat}
\end{equation}
with the normalization appropriately fixed to unity. In the simplest case of a
Gaussian probability distribution, $L(\epslam)$ is a parabola, and
$\sigma_{\epslam}$, the RMS width of $p({\cal D}|\epslam)$, corresponds to the
width of the Gaussian. Even though some of the $L(\epslam)$ functions of the CI
analysis are not parabola--like, $\sigma_{\epslam}$ is still well defined and
can be interpreted as a measure of the experimental sensitivity to a given CI
scenario.  The values of $\sigma_{\epslam}$ are summarized in \tab{reszeus}.

Usually, simple assumptions about $p_\zero(\epslam)$ are made in order to
calculate $p(\epslam|{\cal D})$, which only weakly depends on these assumptions
provided that the width of $p_\zero(\epslam)$ is much larger than
$\sigma_{\epslam}$.  We have calculated $p(\epslam|{\cal D})$ using a flat prior
probability restricted to either $\epsilon\ge0$ or $\epsilon\le0$. One--sided
$95\%$ C.L.\ limits in the Bayesian approach ($\Lambda^\pm_\slimb$) have been
determined by solving\footnote{Similar approaches have been used by ALEPH
\cite{bar-99-01} and CDF \cite{abe-97-03}.}
\begin{equation}
  {\intl_0^{1/(\Lambda^+_\slimb)^2}\;d\xi\,\exp\left(-L(\xi)\right)\over
   \intl_0^\infty\;d\xi\,\exp\left(-L(\xi)\right)}
   =0.95
   \quad\text{and}\quad
  {\intl_{-1/(\Lambda^-_\slimb)^2}^0\;d\xi\,\exp\left(-L(\xi)\right)\over
   \intl_{-\infty}^0\;d\xi\,\exp\left(-L(\xi)\right)}
   =0.95\;;
                                                        \label{eq-baylim}
\end{equation}
For all CI scenarios, the results deviate by less than $20\%$ from the limits
$\Lambda_\slim$ resulting from the MCE method (see \sec{ana}).  Note that
systematic effects have not been considered for this cross check.

\begin{figure}[p]
  \centerline{\psfig{file=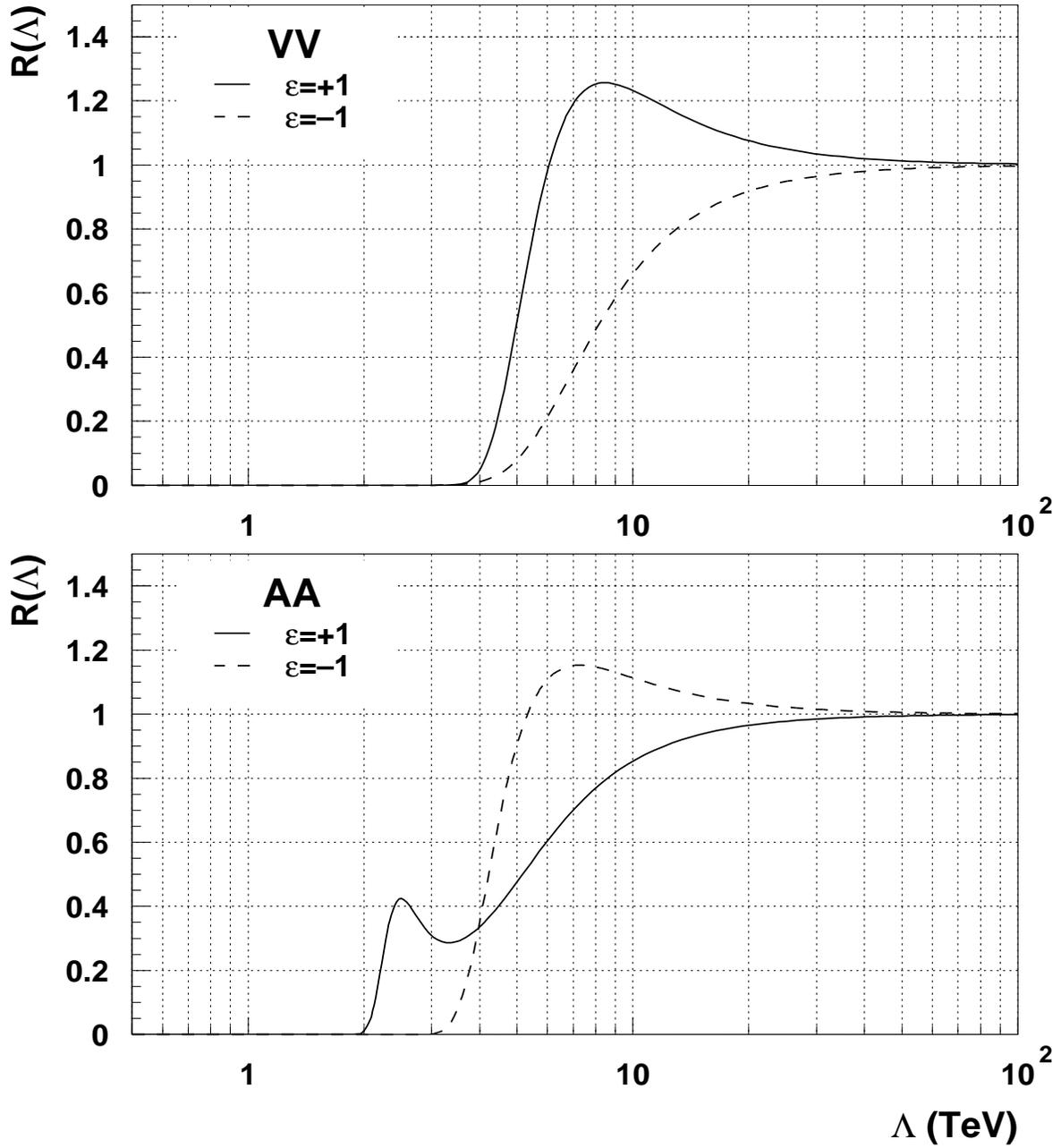,width=15.5cm,height=17.cm}}
  \vspace{-2.mm}
  \caption{The function $R$ defined in \eq{baybur} for the VV (top) and the AA
           (bottom) scenarios, for $\epsilon=+1$ (solid lines), and for
           $\epsilon=-1$ (dashed lines). The graphs are based on the
           polynomial parameterizations of \tab{polco}.}
  \label{fig-burat}
\end{figure}

An alternative way to present the results of this search analysis is obtained by
considering the ratio of two equations of the type (\ref{eq-baypro}) for
different values of $\epslam$, where one is taken as a reference value (chosen
to be $\epslam=0$, i.e.\ corresponding to the SM). Rearranging the terms yields
\begin{equation}
  \left.{p(\epslam|{\cal D})\over p_\zero(\epslam)}\right/
   {p(\epslam=0|{\cal D})\over p_\zero(\epslam=0)}=
   {p({\cal D}|\epslam)\over p({\cal D}|\epslam=0)}=R(\epslam)\;,
                                                        \label{eq-baybur}
\end{equation}
where the double ratio on the left--hand side quantifies how the probability
assigned to a given value of $\epslam$ changes due to the experimental data
${\cal D}$, with the reference point, $\epslam=0$, fixing the normalization.
The function $R(\epslam)$ has been discussed in detail elsewhere (see e.g.\
\cite{dag-99-01} and references therein).  The representation of \eq{baybur} is
independent of the prior probability and can be used to combine the results of
this analysis with those of other experiments by analyzing the product of
corresponding $R$ functions.  Note that $R$ does not involve integrations over
$\epslam$ and is hence invariant with respect to the variable transformations
mentioned in \sec{ana}; in particular, $R$ can be interpreted both as a function
of $\epslam$ and as a function of $\Lambda$.  By definition, $R$ asymptotically
approaches unity if $\epslam\to0$ or $\Lambda\to\infty$, indicating the loss of
experimental sensitivity as the CI strength vanishes. Regions where $R$ is close
to zero are excluded by ${\cal D}$, whereas $R>1$ indicates ``signal--type''
regions where the experimental data are better described by a CI scenario than
by the SM. Typical $R$ values for significant deviations from the SM are
expected to exceed unity by several orders of magnitude (cf.\ the discussion in
\cite{dag-99-01}). Two representative examples of the functions $R(\Lambda)$
observed in the CI analysis are shown in \fig{burat}. In none of the scenarios
from \tab{models} does $R$ exceed unity by more than $40\%$, corroborating the
conclusion that there is no significant indication for the presence of CIs. The
threshold--type region of $R(\Lambda)$ indicates the position of the $\Lambda$
limit. Indeed one obtains limits similar to those reported in \sec{res} by
solving the condition $R(\Lambda^\pm_\slimr)=0.05$ for $\Lambda^\pm_\slimr$.
\vfill\eject

{
\def\bibname{\Large\bf References}
\def\refname{\Large\bf References}
\pagestyle{plain}
\bibliographystyle{./zeusstylet}
{\raggedright
\bibliography{./cipap}
}}

\vfill\eject

\begin{table}[p]
\vbox to \textheight{
\vfill
\vskip 0.5 cm
\renewcommand{\arraystretch}{1.0}
\begin{center}
\begin{tabular}{||c|rrrr|rrrr||}
\hline
Label & $\eta^u_{LL}$ & $\eta^u_{LR}$ & $\eta^u_{RL}$ & $\eta^u_{RR}$ & 
        $\eta^d_{LL}$ & $\eta^d_{LR}$ & $\eta^d_{RL}$ & $\eta^d_{RR}$ \\ 
\hline\hline
VV&  $+1$&$+1$&$+1$&$+1$&
     $+1$&$+1$&$+1$&$+1$ \\
AA&  $+1$&$-1$&$-1$&$+1$&
     $+1$&$-1$&$-1$&$+1$ \\
VA&  $+1$&$-1$&$+1$&$-1$&
     $+1$&$-1$&$+1$&$-1$ \\
X1&  $+1$&$-1$&$ 0$&$ 0$&
     $+1$&$-1$&$ 0$&$ 0$ \\
X2&  $+1$&$ 0$&$+1$&$ 0$&
     $+1$&$ 0$&$+1$&$ 0$ \\
X3&  $+1$&$ 0$&$ 0$&$+1$&
     $+1$&$ 0$&$ 0$&$+1$ \\
X4&  $ 0$&$+1$&$+1$&$ 0$&
     $ 0$&$+1$&$+1$&$ 0$ \\
X5&  $ 0$&$+1$&$ 0$&$+1$&
     $ 0$&$+1$&$ 0$&$+1$ \\
X6&  $ 0$&$ 0$&$+1$&$-1$&
     $ 0$&$ 0$&$+1$&$-1$ \\
U1&  $+1$&$-1$&$ 0$&$ 0$&
     $ 0$&$ 0$&$ 0$&$ 0$ \\
U2&  $+1$&$ 0$&$+1$&$ 0$&
     $ 0$&$ 0$&$ 0$&$ 0$ \\
U3&  $+1$&$ 0$&$ 0$&$+1$&
     $ 0$&$ 0$&$ 0$&$ 0$ \\
U4&  $ 0$&$+1$&$+1$&$ 0$&
     $ 0$&$ 0$&$ 0$&$ 0$ \\
U5&  $ 0$&$+1$&$ 0$&$+1$&
     $ 0$&$ 0$&$ 0$&$ 0$ \\
U6&  $ 0$&$ 0$&$+1$&$-1$&
     $ 0$&$ 0$&$ 0$&$ 0$ \\
\hline
\end{tabular}
\caption{The 30 scenarios for contact interactions considered in this paper.
         Each row of this table corresponds to two different CI scenarios for
         overall interference signs $\epsilon=+1$ and $\epsilon=-1$, 
         respectively (see \eq{cilag} in the text).}
  \label{tab-models}
\end{center}
\vfill
\begin{center}
\begin{tabular}{||c||c|c||}
\hline
quantity & \hbox to 4.cm{\hfil method 1\hfil}
         & \hbox to 4.cm{\hfil method 2\hfil}\\
\hline\hline
$Q^2_\smin\;(\Gev^2)$   &$\cB 500$      &$\cB  400$\\
$Q^2_\smax\;(\Gev^2)$   &$  90200$      &$   51200$\\
\hline
$x_\smin$               &$   0.04$      &\none     \\
$x_\smax$               &$   0.95$      &\none     \\
\hline
$y_\smin$               &$   0.04$      &\none     \\
$y_\smax$               &$   0.95$      &\none     \\
\hline
events                  &$  13243$      &$   37379$\\
\hline
\end{tabular}
\caption{The kinematic regions used for the CI analysis in the two fitting
         methods.  Note that the cuts on $Q^2$, $x$, and $y$ indicated in the
         table are applied in addition to the event selection criteria
         described in \sec{exp}.}
  \label{tab-dasamp}
\end{center}
\vfill}
\end{table}
\begin{table}[p]
  \strut\vfill
  \renewcommand{\arraystretch}{1.2}
  \centerline{
  \begin{tabular}{||c||c|c||c|c||c||}
  \hline
    &\multicolumn{2}{c||}{$\epsilon=-1$}&
     \multicolumn{2}{c||}{$\epsilon=+1$}&\\
  \cline{2-5}
    \hbox to 0.8cm{\hfil \stru{2.8ex}{1.8ex}CI\hfil}
      &\hbox to 2.cm{\hfil $\Lambda^-_\slim$\hfil}
      &\hbox to 2.cm{\hfil $p_\SM$\hfil}
      &\hbox to 2.cm{\hfil $\Lambda^+_\slim$\hfil}
      &\hbox to 2.cm{\hfil $p_\SM$\hfil}
      &\hbox to 2.cm{\hfil $\sigma_{\epslam}$\hfil}\\
    &$(\Tev)$&&$(\Tev)$&&$\Tev^{-2}$\\
  \hline\hline
  VV&  5.0&      & 4.7& 0.28  & 0.021 \\
  \hline
  VA&  2.6& 0.25 & 2.5& 0.25  & 0.070 \\
  \hline
  AA&  3.7& 0.28 & 2.6&       & 0.080 \\
  \hline
  X1&  2.8& 0.26 & 1.8&       & 0.113 \\
  \hline
  X2&  3.1&      & 3.4& 0.28  & 0.056 \\
  \hline
  X3&  2.8&      & 2.9& 0.37  & 0.066 \\
  \hline
  X4&  4.3&      & 4.0& 0.26  & 0.034 \\
  \hline
  X5&  3.3&      & 3.5& 0.28  & 0.052 \\
  \hline
  X6&  1.7& 0.16 & 2.8& 0.27  & 0.105 \\
  \hline
  U1&  2.6& 0.24 & 2.0&       & 0.125 \\
  \hline
  U2&  3.9&      & 4.0& 0.38  & 0.037 \\
  \hline
  U3&  3.5&      & 3.7& 0.48  & 0.046 \\
  \hline
  U4&  4.8&      & 4.4& 0.32  & 0.025 \\
  \hline
  U5&  4.2&      & 4.0& 0.36  & 0.032 \\
  \hline
  U6&  1.8&      & 2.4& 0.24  & 0.118 \\
  \hline
  \end{tabular}}
  \vfill
  \caption{Values of the $95\%$ C.L.\ limits ($\Lambda^\pm_\slim$) for all CI
           scenarios, as well as the SM probabilities, $p_\SM$, for those
           scenarios with $\epsilon/\Lambda_\zero^2\ne0$ (see \sec{ana}).  In
           the last column, the RMS width of the probability distribution
           $p_L\propto\exp(-L(\epslam))$ is indicated (cf.\ Appendix).}
  \label{tab-reszeus}
  \vfill\strut
\end{table}
\begin{table}[p]
  \renewcommand{\arraystretch}{1.2}
  \begin{center}
  \begin{tabular}{||c||c||c|c|c|c|c||}
  \hline
     &\multicolumn{6}{c||}{$\Lambda^\pm_\slim\;(\Tev)$ ($95\%$ C.L.)}\\
  \cline{2-7}
   CI&ZEUS$^a$  &\hbox to 1.6cm{\hfil ALEPH\hfil}
                &\hbox to 1.6cm{\hfil L3\hfil}
                &\hbox to 1.6cm{\hfil OPAL\hfil}
                &\hbox to 1.6cm{\hfil CDF\hfil}
                &\hbox to 1.6cm{\hfil \DO\hfil}\\
     &this      &\cite{bar-99-01}
                &\cite{l3-ci-van,*l3-ci-mor,*l3-ci-dim}
                &\cite{abb-98-01}
                &\cite{abe-97-03}
                &\cite{abb-98-02}\\
     &study&&prelim.&&&\\
  \hline\hline
  VV$+$ & 4.7 & 6.4 & 3.8 & 4.1 & 3.5 & 4.9 \\
  VV$-$ & 5.0 & 7.1 & 5.0 & 5.7 & 5.2 & 6.1 \\
  \hline      		 
  AA$+$ & 2.6 & 7.2 & 5.6 & 6.3 & 3.8 & 4.7 \\
  AA$-$ & 3.7 & 7.9 & 3.5 & 3.8 & 4.8 & 5.5 \\
  \hline      		 
  X1$+$ & 1.8 &\none&\none&\none&\none& 3.9 \\
  X1$-$ & 2.8 &\none&\none&\none&\none& 4.5 \\
  \hline      		 
  X3$+$ & 2.9 & 6.7 & 4.0 & 4.4 &\none& 4.2 \\
  X3$-$ & 2.8 & 7.4 & 3.4 & 3.8 &\none& 5.1 \\
  \hline      		 
  X4$+$ & 4.0 & 2.9 & 2.9 & 3.1 &\none& 3.9 \\
  X4$-$ & 4.3 & 4.5 & 4.8 & 5.5 &\none& 4.4 \\
  \hline      		 
  X6$+$ & 2.8 &\none&\none&\none&\none& 4.0 \\
  X6$-$ & 1.7 &\none&\none&\none&\none& 4.3 \\
  \hline      		 
  U3$+$ & 3.7 &\none& 6.1 & 4.1 &\none&\none\\
  U3$-$ & 3.5 &\none& 4.9 & 5.8 &\none&\none\\
  \hline      		 
  U4$+$ & 4.4 &\none& 2.1 & 2.3 &\none&\none\\
  U4$-$ & 4.8 &\none& 2.9 & 3.2 &\none&\none\\
  \hline
  \end{tabular}
  \begin{minipage}{15.cm}\footnotesize 
  \strut$^a$ No comparison is made for scenarios for which
             only ZEUS sets limits: X2, X5, U1, U2, U5, U6.
  \end{minipage}
  \end{center}
  \caption{Lower $\Lambda$ limits at $95\%$ C.L.\ from this study compared to
           equivalent results from other experiments. The results of the L3
           collaboration are preliminary. The X1, X3/U3, X4/U4, and X6 scenarios
           are denoted LL--LR, LL+RR or V0, LR+RL or A0, and RL--RR,
           respectively, by the LEP and Tevatron experiments. Note that CDF also
           provides limits for $\mu\mu qq$ CI (not shown here) which can be
           combined with the $eeqq$ limits if one assumes lepton flavor
           universality.}
  \label{tab-ciother}
\end{table}
\begin{table}[p]
  \renewcommand{\arraystretch}{1.0}
  \centerline{
  \begin{tabular}{||c||r|r|r|r|r||}
  \hline
  CI       &\multicolumn{1}{c|}{$\lambda_0$}
           &\multicolumn{1}{c|}{$\lambda_1$}
           &\multicolumn{1}{c|}{$\lambda_2$}
           &\multicolumn{1}{c|}{$\lambda_3$}
           &\multicolumn{1}{c||}{$\lambda_4$}\\
  scenario&&\multicolumn{1}{c|}{$\lambda_5$}
           &\multicolumn{1}{c|}{$\lambda_6$}
           &\multicolumn{1}{c|}{$\lambda_7$}
           &\multicolumn{1}{c||}{$\lambda_8$}\\
  \hline\hline
  VV&
  0.219960&--0.317192+2&  0.102541+4&  0.559212+4&--0.157739+5\\
         &&--0.171440+6&  0.296531+6&  0.381829+7&  0.604081+7\\
  \hline
  VA&
  0.245590&  0.137604+1&--0.977350+2&--0.345909+3&  0.131860+5\\
         &&  0.684982+4&--0.160571+6&--0.532035+5&  0.948424+6\\
  \hline
  AA&
  0.241619&  0.136626+2&  0.255357+3&--0.361006+4&  0.477991+4\\
         &&  0.549759+5&--0.700729+5&--0.431092+6&  0.831623+6\\
  \hline
  X1&
  0.206150&  0.700878+1&  0.510132+2&--0.102610+4&  0.197029+4\\
         &&  0.829875+4&--0.124916+5&--0.338700+5&  0.571381+5\\
  \hline
  X2&
  0.198413&--0.113778+2&  0.181979+3&  0.126775+4&  0.163454+4\\
         &&--0.100860+5&--0.147490+5&  0.448248+5&  0.807606+5\\
  \hline
  X3&
  0.082637&--0.608908+1&  0.109116+3&  0.214284+3&  0.131258+3\\
         &&--0.396591+3&--0.110789+3&  0.692834+3&  0.517102+3\\
  \hline
  X4&
  0.327244&--0.213799+2&  0.470532+3&  0.356225+4&  0.558752+3\\
         &&--0.519826+5&--0.454742+5&  0.362723+6&  0.623043+6\\
  \hline
  X5&
  0.258595&--0.142310+2&  0.218532+3&  0.147937+4&  0.816187+3\\
         &&--0.134481+5&--0.960868+4&  0.720345+5&  0.103106+6\\
  \hline
  X6&
  0.143683&--0.486270+1&  0.280939+2&  0.843126+3&  0.235808+4\\
         &&--0.623902+4&--0.141521+5&  0.234297+5&  0.505643+5\\
  \hline
  U1&
  0.277925&  0.822634+1&  0.581735+2&--0.708590+3&  0.828391+3\\
         &&  0.476597+4&--0.472480+4&--0.158164+5&  0.208190+5\\
  \hline
  U2&
  0.086540&--0.100733+2&  0.350929+3&  0.100296+4&  0.118597+3\\
         &&--0.122795+5&--0.957897+4&  0.960528+5&  0.159882+6\\
  \hline
  U3&
  0.016648&--0.397315+1&  0.241440+3&  0.103296+3&--0.139481+3\\
         &&--0.115302+2&  0.154464+3&--0.851289+2&  0.437650+3\\
  \hline
  U4&
  0.145224&--0.224785+2&  0.824204+3&  0.231869+4&--0.154694+5\\
         &&--0.237649+5&  0.454907+6&  0.537245+5&--0.486230+7\\
  \hline
  U5&
  0.108391&--0.149704+2&  0.498695+3&  0.981530+3&--0.449846+4\\
         &&--0.546963+4&  0.877442+5&  0.215059+3&--0.652759+6\\
  \hline
  U6&
  0.174813&--0.417751+1&  0.981713+1&  0.462709+3&  0.134709+4\\
         &&--0.284097+4&--0.650313+4&  0.835725+4&  0.172630+5\\
  \hline
  \end{tabular}}
  \caption{Parameterizations of the functions $L(\epslam)-L_\smin$ resulting
           from the unbinned method. For each CI scenario, the nine coefficients
           $\lambda_i$ define a polynomial $\sum_{i=0}^8\lambda_i(\epslam)^i$
           which has been fitted to $L(\epslam)-L_\smin$ in the range where 
           $L(\epslam)-L_\smin<18$. The notation $x\pls n$ is used as a 
           shorthand for $x\cdot10^n$.}
  \label{tab-polco}
  \renewcommand{\arraystretch}{1.1}
\end{table}

\begin{figure}[p]
  \centerline{\psfig{file=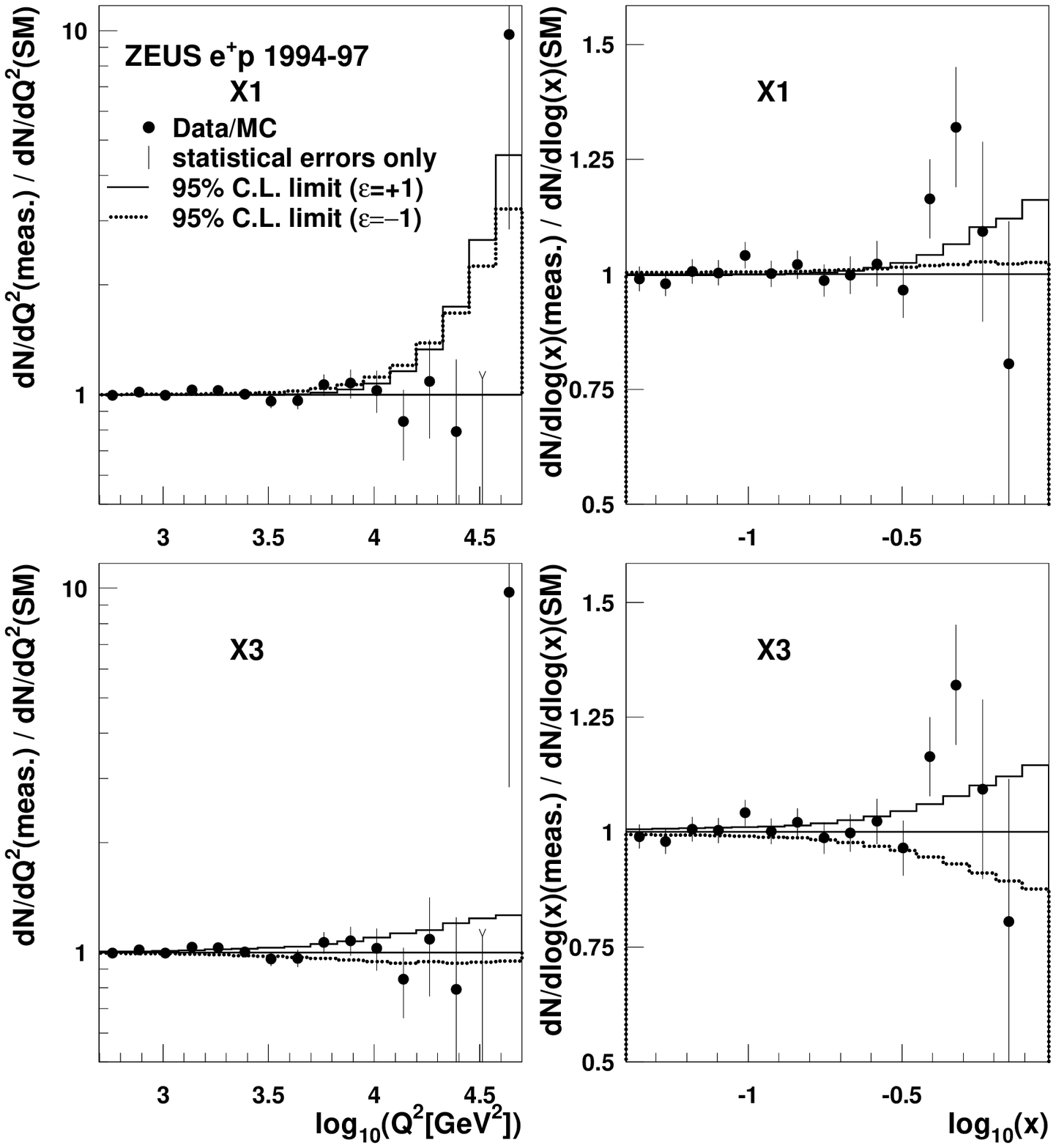,width=\textwidth,height=19.cm}}
  \vspace{-2.mm}
  \caption{Exclusion limits on CI strengths in terms of the corresponding
           modification of the expected distributions of $Q^2$ (left) and $x$
           (right) in the kinematic region used for method~1 (see \sec{ana}
           and \tab{dasamp}).  The dots represent the ratios of observed and
           expected numbers of events and the error bars indicate the
           statistical uncertainties. The histograms show the modification of
           this ratio for the X1 (top) and X3 (bottom) scenarios as obtained
           with the $\Lambda_{\rm lim}$ values given in \tab{reszeus}. The
           solid (dotted) line corresponds to the $95\%$ exclusion limit for
           $\epsilon=+1$ ($\epsilon=-1$).}
  \label{fig-xqdisb}
\end{figure}

%
%
\end{document}